\documentclass[pra,superscriptaddress,aps,longbibliography,twocolumn]{revtex4-1}%
\usepackage{amsfonts}
\usepackage{mathtools}
\usepackage{enumerate}
\usepackage{amsmath}
\usepackage{amssymb}
\usepackage{pgfplots, pgfplotstable}
\usepackage{dsfont}
\usepackage{tikz}
\usepackage{graphicx}
\usepackage{xcolor}
\usepackage{amsmath}%
\usepackage{amsfonts}%
\usepackage[T1]{fontenc}
\usepackage{amssymb}%
\usepackage{graphicx}
\usepackage{enumitem}
\usepackage{physics}
\usepackage{appendix}
\usepackage{subfigure}
\usepackage{thmtools,thm-restate}

\newtheorem{theorem}{Theorem}

\newtheorem{corollary}[theorem]{Corollary}

\newtheorem{proposition}[theorem]{Proposition}

\newenvironment{proof}[1][Proof]{\noindent\textbf{#1.} }{\ \rule{0.5em}{0.5em}}

\usepackage[colorlinks=true,linkcolor=blue,citecolor=red,plainpages=false,pdfpagelabels]
 {hyperref}
\usepackage{cleveref}
\usepackage{color}
\definecolor{dgreen}{HTML}{006600}

\allowdisplaybreaks

\begin{document}
\preprint{ }
\title[]{Asymptotic security of discrete-modulation protocols for continuous-variable\\quantum key distribution}
\author{Eneet Kaur}
\affiliation{Hearne Institute for Theoretical Physics, Department of Physics and Astronomy, and Center for Computation and Technology,
Louisiana State University, Baton Rouge, Louisiana 70803, USA}
\author{Saikat Guha}
\affiliation{College of Optical Sciences and Department of Electrical and Computer Engineering, University of Arizona,
 Tucson, Arizona 85719, USA}
\author{Mark M. Wilde}
\affiliation{Hearne Institute for Theoretical Physics, Department of Physics and Astronomy, and Center for Computation and Technology,
Louisiana State University, Baton Rouge, Louisiana 70803, USA}

\begin{abstract}
We consider discrete-modulation protocols for continuous-variable quantum key distribution (CV-QKD) that employ a modulation constellation consisting of a finite number of coherent states and that use a homodyne or a heterodyne-detection receiver. We establish a security proof for collective attacks in the asymptotic regime, and we provide a formula for an achievable secret-key rate. Previous works established security proofs for discrete-modulation CV-QKD protocols that use two or three coherent states. The main constituents of our approach include approximating a complex, isotropic Gaussian probability distribution by a finite-size Gauss-Hermite constellation, applying entropic continuity bounds, and leveraging previous security proofs for Gaussian-modulation protocols. As an application of our method, we calculate secret-key rates achievable over a lossy thermal bosonic channel. We show that the rates for discrete-modulation protocols approach the rates achieved by a Gaussian-modulation protocol as the constellation size is increased. For pure-loss channels, our results indicate that in the high-loss regime and for sufficiently large constellation size, the achievable key rates scale optimally, i.e., proportional to the channel's transmissivity.

\end{abstract}
\volumeyear{year}
\volumenumber{number}
\issuenumber{number}
\eid{identifier}
\date{\today}
\startpage{1}
\endpage{10}
\maketitle

\section{Introduction}

Quantum key distribution (QKD) allows for two distant parties, often called Alice and Bob, to create a shared secret key by employing an insecure and noisy quantum communication channel and an authenticated public classical communication channel \cite{Bennett84,E91,Scarani2009}. The security is based on the physical laws of quantum mechanics, in contrast to conventional cryptographic protocols, whose security relies on computational complexity-theoretic assumptions. 

There are two basic classes of QKD protocols that have been considered: discrete-variable and continuous-variable  (see, e.g., \cite{Scarani2009} for a review). In discrete-variable QKD (DV-QKD), the information is usually encoded in the polarization or time bin  of single photons or weak coherent states (laser-light pulses). Discrete-variable QKD requires high-efficiency, low dark-count-rate, single-photon detectors, which are expensive and often need extreme cryo-cooling. In the other class of protocols, known as continuous-variable QKD (CV-QKD), the information is encoded in the quadrature amplitudes of coherent states.
The transmitter modulates the phase and/or the amplitude of laser-light pulses, and the receiver is based on coherent detection (i.e., homodyne or heterodyne detection). Near shot-noise-limited, low-noise homodyne/heterodyne detection is readily realizable at room temperature using off-the-shelf hardware, unlike the single-photon detectors of DV-QKD.
CV-QKD protocols thus possess a major advantage over DV-QKD in terms of the cost and ease of experimental implementation.

However, one major area that DV-QKD currently possesses an advantage over CV-QKD is that the DV modulation involves few levels (e.g., two polarization states of a photon or three amplitude levels of a coherent state in the decoy-state BB84 protocol \cite{H03,W05,LMC05}), which puts far less burden on the transmitter's modulator compared to that of the traditional Gaussian-modulation CV-QKD protocol. The latter  requires modulation using an infinite-size constellation.
This also makes the error correction protocols far simpler for DV-QKD, along with much less overhead for random-number generation. Another area where DV-QKD is arguably more advanced is the availability of quantum repeater protocols
\cite{JTNMVL09,MLKLLJ16,GKFDSST15,PKEG17,MZLJ18} for overcoming the fundamental rate-vs.-loss trade-off of direct-transmission based QKD \cite{TGW14,PLOB17,WTB16}. However, there have been recent advances in designs of repeaters for CV-QKD \cite{DR17,FM18,SKG18}. For experimental developments in CV-QKD see \cite{DinhXuan2009, Wang2013, Huang2015, Wang2019, MDI-QKD-exp, one-sided-cv-exp}.  

In the most common form of CV-QKD, one uses Gaussian modulation of coherent states \cite{GG02}: Alice modulates laser-light pulses with amplitudes selected randomly from a complex-valued Gaussian distribution with a given variance. Security proofs for this Gaussian modulation CV-QKD protocol have been developed for arbitrary attacks, even in the finite key-length regime \cite{L15}. Additionally, a suite of variants of this CV-QKD protocol exist, some of which use squeezed light modulation and two-way transmission \cite{CLA01,FFBLSTW12,WLBSRL04,LGRC13,UG15,UF10,WPLR10,FC12,WRSL13}.

However, all of their asymptotic security proofs require a Gaussian modulation. Gaussian modulation has obvious drawbacks, which include extreme burden on the transmitter's random number source, as well as computationally demanding and inefficient error-correction techniques. Furthermore, no matter how high the extinction ratio of a practically-realizable electro-optic modulator, it is impossible to sample pulse amplitudes from a true Gaussian distribution, on which the security proofs rely.

Despite the fact that Gaussian modulation has made security proofs manageable, it is important---for the practical realizability of CV-QKD---that protocols that use a few pre-determined modulation levels (such as binary phase and quadrature amplitude modulation) are proven secure. 
Discrete-modulation CV-QKD was introduced in \cite{Ralph99,PhysRevA.62.062308,HYAKN03}, where the coherent states transmitted in each mode are chosen according to a discrete probability distribution, and it was developed further in \cite{Zhao2009}. Discrete-modulation CV-QKD protocols can leverage the efficient modulation and error correction, and low-overhead random number generation that DV-QKD enjoys, while retaining the ease of implementation of homodyne/heterodyne detection of CV-QKD.

Several discrete-modulation protocols have already been considered \cite{Zhao2009,Bradler2018,Leverrier1,Leverrier2}, and security proofs have been developed in the asymptotic regime, i.e., in the limit of a large number of  uses of the quantum channel, hence generating a large-length key (at a given key-bits per channel-use rate). Ref.~\cite{Zhao2009} considered a protocol with binary-phase shift-keying  of coherent states along with homodyne detection. However, the secure key rate established there is more than an order of magnitude lower than that which can be achieved with Gaussian modulation. Motivated by \cite{Zhao2009}, Ref.~\cite{Bradler2018} considered ternary-phase shift-keying  modulation with homodyne detection, which led to an improvement in the secure key rates, but the resulting secret-key rates are still far from the key rates achievable with Gaussian modulation. Refs.~\cite{Leverrier2, Leverrier1} established security for discrete-modulation protocols against particular collective attacks that correspond to linear bosonic channels. For other protocols that use discrete modulation of coherent states, see \cite{use-defined-qkd, Sych2010}. 

This brings us to the long-standing open problem of proving security of 
a general $M$-ary discrete-modulation CV-QKD protocol, for $M$ beyond a minimum threshold value, with the feature that the achievable key rate approaches that of Gaussian modulation as $M$ goes to infinity. Such a result is of significant value for the practical usability of CV-QKD. In this paper, we accomplish the aforesaid for security against  collective attacks having the physically reasonable assumptions outlined in Section~\ref{sec:channel-assumptions}. Establishing a security proof and key-rate lower bounds for discrete modulation CV-QKD protocols with a finite key length is left open for future work. Our proof eliminates the need to consider protocols based on Gaussian modulation in order to have asymptotic security in CV-QKD, with the ability of the user to determine the size of the modulation alphabet based on how close one desires the key rates to be to the Gaussian modulation protocol. In addition, our numerical evaluation of achievable key rates over a pure-loss bosonic channel suggests that, for sufficiently large constellation size, the achievable key rates are proportional to the channel's transmissivity, which is known to be the optimal rate-vs.-loss scaling achievable with any QKD protocol, CV or DV~\cite{TGW14}.

To establish these results, we make use of two important recent theoretical advances: the approximation of Gaussian distributions with discrete ones for communication \cite{Verdu,Renes}, especially in the context of bosonic Gaussian states \cite{Renes}, and an entropic continuity bound from \cite{Shirokov} for energy-bounded bosonic states. 
The idea of approximating a Gaussian modulation with a discrete one for CV-QKD was proposed in \cite{JKDL12}, but this work did not provide a security proof for CV-QKD with discrete modulation. One of the main tools, beyond the approaches considered in \cite{JKDL12} and which allows us to establish a security proof, is the entropic continuity bound from \cite{Shirokov}. We also develop methods for using the parameters observed in a discrete-modulation CV-QKD protocol to bound Eve's Holevo information. 

This paper is organized as follows. We introduce discrete-modulation CV-QKD in Section~\ref{sec:protocol}, followed by Section~\ref{sec:channel-assumptions}'s detailed list of our assumptions on the collective attack of an eavesdropper. We give our security proof in Section~\ref{sec:proof}, and we discuss details of channel estimation in Section~\ref{sec:channel_estimation}. We then showcase, in Section~\ref{sec:numerics}, the secure key rates that our approach leads to when the protocol is conducted over a lossy thermal bosonic channel. We end with open questions and future directions in Section~\ref{sec:conclusion}.

\textit{Note}: In work independent of and concurrent to ours, other approaches for security proofs in discrete modulation of CV-QKD have been put forward \cite{PhysRevX.9.021059,PhysRevX.9.041064}.

\section{Protocol}

\label{sec:protocol}

We begin by outlining the steps of a phase-symmetrized discrete-modulation CV-QKD protocol based on  $m^2$ coherent states, where $m \in \mathbb{N}$. In this protocol, Bob  performs either homodyne or heterodyne detection. Let $X$ be a random variable with realizations $x\in \{1,2,\ldots, m^2\}$ and fix $\alpha_x \in \mathbb{C}$ for all $x$. Let $r(x)$ be the probability associated with the realization $x$. The steps of the protocol are as follows:
 
\begin{enumerate}
    \item Alice prepares the coherent state $\ket{\alpha_x}$ with probability $r(x)$.
    She records the value of $x$ in the variable $x_j$, where $j \in \{1,\ldots, n\}$ refers to the transmission round. 
    She also records the value $\sqrt{2} \Re{\alpha_x}$ in the variable~$q_j$ and the value $\sqrt{2} \Im{\alpha_x}$ in the variable~$p_j$ . 
    Exact expressions for $\alpha_x$ and $r(x)$ that we use in the protocol are given in Section~\ref{sec:channel_estimation}. 
    
    \item 
    Alice then picks a phase $\phi_j \in \left\{0, \pi/2, \pi, 3\pi / 2 \right\}$ uniformly at random, applies it to her channel input mode as the unitary $e^{-i \hat{n} \phi_j}$, which is physically realized by a phase shifter. The resulting state is then  $e^{-i \hat{n} \phi_j}\ket{\alpha_x} = \ket{\alpha_x e^{-i  \phi_j}}$, which she transmits over the unknown and insecure quantum communication channel $\mathcal{N}$ to Bob. At the same time, she communicates the choice $\phi_j$ to Bob over a public authenticated classical channel and then she locally discards or forgets the choice of $\phi_j$.
    The insecure quantum channel $\mathcal{N}$ can be controlled by an eavesdropper Eve. Our assumptions on the insecure quantum channel
    $\mathcal{N}$
    are stated in Section~\ref{sec:channel-assumptions}.

  \item Upon receiving the output of the quantum channel, namely, the state $\mathcal{N}(e^{-i \hat{n} \phi_j} |\alpha_x\rangle\langle \alpha_x |e^{i \hat{n} \phi_j})$, as well as the classical choice of $\phi_j$ from the public authenticated classical channel,
  Bob applies the reverse phase as the inverse unitary $e^{i \hat{n} \phi_j}$, and then locally discards or forgets the value of $\phi_j$. The resulting state is then as follows:
  \begin{equation}
      \overline{\mathcal{N}}(|\alpha_x\rangle \langle \alpha_x|),
  \end{equation}
  where the phase-symmetrized channel $\overline{\mathcal{N}}$ is defined as
  \begin{equation}
\overline{\mathcal{N}}(\rho)\equiv 
\frac{1}{4}\sum_{k=0}%
^{3}U(k)^\dag \mathcal{N}(U(k)\rho U(k)^\dag)U(k),
\label{eq:phase-sym-channel}
\end{equation}
with $U(k) \equiv e^{-i\hat{n} \pi
k/2}$. The phase symmetrization of the channel $\mathcal{N}$  is helpful in reducing the number of parameters that need to be estimated during the channel estimation part of the protocol, as we explain in Section~\ref{sec:channel_estimation}.
  
\item  If Bob performs position-quadrature or real-quadrature homodyne detection on the state $\overline{\mathcal{N}}(|\alpha_x\rangle \langle \alpha_x|)$ the result is recorded in the variable~$y_j^q$ \footnote{The phase symmetrization performed in Steps 2-3 of the protocol implies that Bob need only perform measurements in one quadrature. That is, the effect of phase symmetrization is to symmetrize Eve's attack evenly with respect to both the position and momentum quadratures, and therefore a measurement of only one of the quadratures suffices to detect Eve's tampering. It is for this reason that we have elected to simplify the CV-QKD protocol so that all of the homodyne measurements are conducted with respect to a single quadrature.}. If Bob performs heterodyne detection, then the value of the position quadrature is recorded in~$y_j^q$, and the value of the momentum quadrature is recorded in $y_j^p$.  
  
  \item Steps 1-4 are repeated $n$ times, for $n$ a large positive integer. If Bob performs homodyne detection, then the sequence $\{q_j\}_{j=1}^n$ is known to Alice, and the sequence $\{y_j^q\}_{j=1}^n$ is known to Bob. If Bob performs heterodyne detection, then the sequences $\{q_j\}_{j=1}^n$ and $\{p_j\}_{j=1}^n$ are known to Alice, and $\{y_j^q\}_{j=1}^n$ and $\{y_j^p\}_{j=1}^n$ are known to Bob. 
  
  \item A constant fraction $\delta$ of the rounds are used for channel estimation (or parameter estimation), for $\delta \in (0,1)$ a small number. That is, for these $\delta n$ rounds, the parameters $\gamma_{11}$, $\gamma_{22}$, and $\gamma_{12}$ are calculated. If Bob performs homodyne detection, then these parameters are given as
  \begin{align}
   \gamma_{11} & \equiv \frac{1}{\delta n} \sum_{j=1}^{\delta n} (q_j - \overline{q})^2   ,\label{eq:gamma11-param} \\
   \gamma_{12} & \equiv \frac{1}{\delta n} \sum_{j=1}^{\delta n} (q_j - \overline{q})( y_j^q - \overline{y}),\\
   \gamma_{22} & \equiv \frac{1}{\delta n} \sum_{j=1}^{\delta n} (y_j^q - \overline{y})^2,
   \label{eq:gamma22-param}
  \end{align}
  where 
  \begin{equation}
      \overline{q}\equiv \frac{1}{\delta n} \sum_{j=1}^{\delta n} q_j, \qquad 
      \overline{y}\equiv \frac{1}{\delta n} \sum_{j=1}^{\delta n} y_j^q.
  \end{equation}
  If Bob performs heterodyne detection, then these parameters are given as
  \begin{align}
   \gamma_{11} & \equiv \frac{1}{\delta n} \sum_{j=1}^{\delta n} (q_j - \overline{q})^2  = \frac{1}{\delta n} \sum_{j=1}^{\delta n} (p_j - \overline{p})^2   ,\label{eq:gamma11-param-hetero} \\
   \gamma_{12} & \equiv \frac{1}{2\delta n} \sum_{j=1}^{\delta n} (q_j - \overline{q})( y_j^q - \overline{y}^q)  +  (p_j - \overline{p})( y_j^p - \overline{y}^p),\\
   \gamma_{22} & \equiv \frac{1}{2\delta n} \sum_{j=1}^{\delta n} (y_j^q - \overline{y}^q)^2+  (y_j^p - \overline{y}^p)^2,
   \label{eq:gamma22-param-hetero}
  \end{align}
  where 
  \begin{align}
      \overline{q}\equiv \frac{1}{\delta n} \sum_{j=1}^{\delta n} q_j, \qquad 
      \overline{y}^q\equiv \frac{1}{\delta n} \sum_{j=1}^{\delta n} y_j^q,\\
      \overline{p}\equiv \frac{1}{\delta n} \sum_{j=1}^{\delta n} p_j, \qquad 
      \overline{y}^p\equiv \frac{1}{\delta n} \sum_{j=1}^{\delta n} y_j^p.
  \end{align}

  Clearly, the parameter $\gamma_{11}$ can be calculated from Alice's data alone, $\gamma_{22}$ can be calculated from Bob's data alone, but it is necessary to calculate $\gamma_{12}$ from both Alice and Bob's data, and so it is necessary for Bob to share the $y_j$ values of these $\delta n$ rounds with Alice over a public authenticated classical channel. Furthermore, the public authenticated classical channel is used for Alice and Bob to share the values of $\gamma_{11}$, $\gamma_{12}$, and $\gamma_{22}$ with each other. The data $x_j$, $q_j$, $p_j$ and $y_j^{q,p}$ for these $\delta n$ channel estimation rounds are then discarded. A detailed analysis of the channel estimation part of the protocol is given in Section~\ref{sec:channel_estimation}.
  
    \item The remaining $q_j$, $p_j$, and $y_j^{q,p}$ data are used for final key generation. The final key-generation protocol includes reverse reconciliation, error correction, and privacy amplification (see \cite{Scarani2009} for a review). 
\end{enumerate}

\section{Assumptions on the insecure quantum communication channel}

\label{sec:channel-assumptions}

In this section, we outline the various assumptions that we make on the insecure quantum communication channel:

\begin{enumerate}

\item 
Each Alice-to-Bob transmission is assumed to take place over independent identical uses of a quantum channel ${\cal N}$, which is unknown to Alice and Bob at the beginning of the protocol. We assume that any deviation of ${\cal N}$ from the identity channel is attributed to the most general adversarial action by Eve. Even though Eve's action---which appears as a noisy quantum channel ${\cal N}$ to Alice and Bob---remains the same for each transmission, she is allowed to make arbitrary collective measurements on her quantum system at the end of the protocol. See below for a mathematical description. This scenario is referred to as a {\em collective attack}.

\item The channel is described mathematically as an isometric quantum channel $\mathcal{U}_{A \to BE}$, meaning that there exists an isometry $U_{A\to BE}$, satisfying $[U_{A\to BE}]^\dag U_{A\to BE}=I_A$, such that
\begin{equation}
\mathcal{U}_{A \to BE}(\rho_A) \equiv U_{A\to BE}\, \rho_A\, (U_{A\to BE})^\dag
\end{equation}
for all input density operators $\rho_A$. The systems $A$, $B$, and $E$ are described by separable Hilbert spaces $\mathcal{H}_A$, $\mathcal{H}_B$, and $\mathcal{H}_E$, respectively. The system $A$ corresponds to a single bosonic mode, and system $B$ does also. In particular, the channel can accept coherent states at the input $A$ and is such that the receiver can perform homodyne or heterodyne detection on the system $B$. The system $A$ is accessible to the sender Alice, the system $B$ is accessible to the receiver Bob, and the system $E$ is in possession of the eavesdropper Eve. 

\item The reduced channel from Alice to Bob is given by
\begin{equation}
\mathcal{N}_{A \to B}(\rho_A) \equiv \operatorname{Tr}_E[\mathcal{U}_{A \to BE}(\rho_A)],
\label{eq:reduced-channel}
\end{equation}
and this channel $\mathcal{N}_{A \to B}$ is what is used in the protocol description in Section~\ref{sec:protocol}.
We assume that if the mean photon number of the input state $\rho_A$ is finite, then the mean photon number of the output state $\mathcal{N}_{A \to B}(\rho_A)$ is finite. That is, $\operatorname{Tr}[\hat{n}\mathcal{N}_{A \to B}(\rho_A)] < \infty$ if $\operatorname{Tr}[\hat{n}\rho_A] < \infty$. Furthermore, we assume that if the variance of the photon number of the input state $\rho_A$ is finite, then the variance of the photon number of the output state $\mathcal{N}_{A \to B}(\rho_A)$ is finite. This implies that $\operatorname{Tr}[\hat{n}^2\mathcal{N}_{A \to B}(\rho_A)] < \infty$ if $\operatorname{Tr}[\hat{n}^2\rho_A] < \infty$.

\item We assume that if the mean photon number of the input state $\rho_A$ is finite, then the mean energy of Eve's state $\operatorname{Tr}_B[\mathcal{U}_{A \to BE}(\rho_A)]$ is finite, where the mean energy is computed with respect to a physically reasonable Hamiltonian $H_E$ that satisfies the Gibbs hypothesis \cite{Holevo03,Holevo04,Winter}, meaning that $\operatorname{Tr}[e^{-\beta H_E}] < \infty$ for all $\beta > 0$ and has its ground-state energy equal to zero. For example, if Eve's system $E$ of the state $\operatorname{Tr}_B[\mathcal{U}_{A \to BE}(\rho_A)]$ consists of several bosonic modes $E_1, \ldots, E_k$, then $H_E$ could be taken as the total photon number operator $\hat{n}_1 + \cdots + \hat{n}_k$ for all of  the $k$ modes. 

\item Let
\begin{equation}
 \mu(q_A,p_A) \equiv \int dq_B\, r_{Q_B|Q_A,P_A}(q_B|q_A p_A)\, q_B   ,
\end{equation}
denote the conditional mean of the position quadrature of Bob, where $r_{Q_B|Q_A,P_A}(q_B|q_A p_A)$ is the conditional probability distribution of the position quadrature $q_B$ of the state \begin{equation}
    \sigma_{B}^{q_A,p_A} \equiv \mathcal{N}_{A\rightarrow B}(\op{\alpha(q_A,p_A)}_A),
\end{equation}
and $\op{\alpha(q_A,p_A)}$ is a coherent state with position quadrature $q_A$ and momentum quadrature $p_A$. We suppose that $\mu(q_A,p_A)=\sum_{k=0}^{K_1}\sum_{l=0}^{K_2} \mu_{kl} q_A^kp_A^l$, where $K_1,K_2 \in \mathbb{Z}^+$. That is, the mean value of the position quadrature of $\sigma_B^{q_A,p_A}$ is no more than a $K_1\textrm{th}$-order polynomial in $q_A$ and a $K_2\textrm{th}$-order polynomial in $p_A$. We also suppose that $\mu_{kl}$ is an exponentially decaying function, $\exp[-a(k+l)]$, in $k$ and $l$ for $k\geq 2m-2$ and $l\geq 2m-1$. Here, $a> 0$ and $m$ is the constellation size. For simplicity, we suppose that $K_1=K_2$. These assumptions are required for the security proof presented in Appendix~\ref{app:estimategaussian}. 
\end{enumerate}

We note that an immediate consequence of the bounded mean photon number assumption in part three above, by applying the Cauchy--Schwarz inequality, is the following: If Alice inputs a state $\rho_A$ with finite mean vector $[\langle \hat{q}\rangle_\rho, \langle \hat{p}\rangle_\rho]$, then the output mean vector for the state of system $B$ is finite. If the input state $\rho_A$ has a finite covariance matrix with entries given by
\begin{equation}
\begin{bmatrix}
2 \langle \hat{q}_0^2 \rangle_{\rho} &
\langle \hat{q}_0 \hat{p}_0 + \hat{p}_0 \hat{q}_0 \rangle_{\rho}
\\
\langle \hat{q}_0 \hat{p}_0 + \hat{p}_0 \hat{q}_0 \rangle_{\rho} & 2 \langle \hat{p}_0^2 \rangle_{\rho}\\
\end{bmatrix} ,   
\end{equation}
where $\hat{q}_0 \equiv \hat{q} - \langle \hat{q}\rangle_\rho$ and $\hat{p}_0 \equiv \hat{p} - \langle \hat{p}\rangle_\rho$,
then the covariance matrix of the output state
$\mathcal{N}_{A \to B}(\rho_A)$
is finite.

\section{Secret-key rate lower bound}


\label{sec:proof}

The asymptotic secret-key rate $K$ is bounded from below by the Devetak-Winter formula \cite{Devetak2005,Kraus2005} as 
\begin{equation}
    K \geq I(X;Y)-\sup_{\mathcal{U}_{A\to BE} \in \mathcal{S}} \chi(Y;E).
    \label{eq:DW-formula}
\end{equation}
In the inequality above, the Shannon mutual information between Alice's variable $X$ and Bob's variable $Y$ is denoted by $I(X;Y)$, and the Holevo information between Bob's variable $Y$ and Eve's quantum system $E$ is denoted by $\chi(Y;E)$. We suppose that the quantum channel connecting Alice to Bob is not known, satisfies the assumptions given in Section~\ref{sec:channel-assumptions}, and can only be partially estimated from $X$ and the measurement outcomes $Y$ on Bob's side, as we discuss in Section~\ref{sec:channel_estimation}. This lack of knowledge is an advantage to Eve. Therefore, the inequality in \eqref{eq:DW-formula} features an optimization of the Holevo information $\chi(Y;E)$ over all isometric quantum channels $\mathcal{U}_{A\to BE}$  of Eve that are compatible with Alice's and Bob's data. Let~$\mathcal{S}$ denote the set of channels that are consistent with the measurement data. We discuss the precise meaning of this statement in  Section~\ref{sec:channel_estimation}.  We also suppose that reverse reconciliation \cite{Grangier2002} is being used in the key-generation protocol, in which the public classical communication is from Bob to Alice, and this accounts for Bob's variable $Y$ appearing in the $\chi(Y;E)$ term in~\eqref{eq:DW-formula}.  

To calculate  the lower bound in \eqref{eq:DW-formula}, we first need to calculate the Shannon mutual information $I(X;Y)$, which can be easily obtained from the observed data of Alice and Bob. The main difficulty is then to perform the optimization over the isometric quantum channels $\mathcal{U}_{A\to BE}$ of Eve and to bound the Holevo information $\chi(Y;E)$ from above. Doing so is the main bottleneck for many security proofs in quantum key distribution. 

For protocols involving Gaussian modulation of coherent states, the aforementioned problem was solved in \cite{Patron2006, Navascus}, with \cite{Patron2006} relying on the techniques of \cite{Wolf}. The optimal attack by Eve for such protocols was proved to be a Gaussian attack, which considerably simplifies the security analysis. However, once we consider discrete-modulation protocols, the optimal attack by Eve is no longer known, and is unlikely to be Gaussian. To address this problem, novel techniques are required.

In this paper, we provide a security proof for the protocol described in Section~\ref{sec:protocol} by employing  various existing tools:  the approximation of Gaussian distributions with discrete ones \cite{Verdu,Renes}, an entropic continuity bound from \cite{Shirokov}, and  the optimality of Gaussian attacks for Gaussian modulation of coherent states \cite{Patron2006, Navascus}. The approach that we employ in this paper is rather intuitive: we approximate the Gaussian distribution with a discrete distribution and bound the error introduced due to this approximation in trace norm, by employing the techniques of \cite{Verdu,Renes}. Then, we expect Eve's Holevo information due to this approximation to be close to Eve's Holevo information resulting from a Gaussian-modulated protocol, with the absolute value of the difference being a function of the error introduced in the approximation.

We now discuss this approach in detail. First, consider a key-generation protocol that employs coherent states with Gaussian modulation. The expected density operator for Alice's transmitted state is a thermal state $\theta(N_S)$ with mean photon number $N_S \geq 0$:
\begin{equation}
    \theta(N_S) \equiv \frac{1}{N_S+1} \sum_{n=0}^\infty \left(\frac{N_S}{N_S+1}\right)^n |n\rangle\langle n|.
\end{equation}
The $P$-function of the thermal state $\theta(N_S)$ is a circularly symmetric complex Gaussian \cite{GK04}. Following the approach of \cite{Renes}, we can approximate the real and imaginary parts of the circularly symmetric Gaussian by the various constellations considered in \cite{Verdu}: Gauss-Hermite, random walk, equilattice, and quantile. The type of constellation fixes $\ket{\alpha_x}$ and $r(x)$. In this paper, we focus exclusively on the Gauss-Hermite constellation. It is possible to consider other constellations and obtain security proofs for these other constellations using the techniques described below. We obtain the error introduced by this approximation, by employing bounds from \cite{Renes},  and then we apply an entropic continuity bound from \cite{Shirokov} to obtain an upper bound on Eve's Holevo information $\chi(Y;E)$. 

We now discuss our security proof for discrete-modulation protocols of the form presented in Section~\ref{sec:protocol}. Suppose that Alice employs the following discrete-modulation
ensemble of coherent states:%
\begin{equation}
\{r(x),|\alpha_{x}\rangle\}_{x=1}^{m^2},
\end{equation}
with expected density operator:%
\begin{equation}
\overline{\rho}\equiv\sum_{x=1}^{m^2} 
r(x)|\alpha_{x}\rangle\langle \alpha_{x}|.
\end{equation}
Then depending on the constellation size $m^2$ and the mean photon number $N_S$ of the thermal state being approximated, we obtain the following bound on the normalized trace distance:
\begin{equation}
\frac{1}{2}\left\Vert \overline{\rho}-\theta(N_{S})\right\Vert _{1}%
\leq\varepsilon(m,N_S), \label{eq:error}
\end{equation}
where $\theta(N_{S})$ is a thermal state of mean photon number $N_{S}$ and $\varepsilon(m,N_S)$ is the approximation error, for which we determine an explicit characterization later  in \eqref{eq:error1}, by employing the techniques of \cite{Renes}.

The secret-key rate with reverse reconciliation is given by%
\begin{equation}
\beta I(X;Y)-\chi(Y;E),
\end{equation}
where $\beta$ is the reconciliation efficiency \cite{PhysRevA.76.042305} and the mutual information quantities are computed with respect to the
following ensemble:%
\begin{equation}
\{r(x,y),\rho_{E}^{x,y}\}_{x,y},
\end{equation}
where%
\begin{align}
r(x,y)  & \equiv r(x)r(y|x),  \\
r(y|x)  & \equiv \operatorname{Tr}\{(\Lambda_{B}^{y}\otimes I_{E})\mathcal{U}%
_{A\rightarrow BE}(|\alpha_{x}\rangle\langle\alpha_{x}|_A)\},  \\
\rho_{E}^{x,y}  & \equiv \frac{1}{r(y|x)}\operatorname{Tr}_{B}\{(\Lambda_{B}%
^{y}\otimes I_{E})\mathcal{U}_{A\rightarrow BE}(|\alpha_{x}\rangle
\langle\alpha_{x}|_A)\},
\end{align}
with $\{\Lambda^{y}\}_{y}$ denoting Bob's POVM and $\mathcal{U}_{A\rightarrow
BE}$ the isometric channel satisfying the assumptions of Section~\ref{sec:channel-assumptions} and corresponding to the collective attack of Eve. Since we do
not know what collective attack Eve will employ, we minimize the secret-key
rate with respect to all collective attacks that are consistent with the
measurement data observed by Alice and Bob, i.e., with respect to all isometric channels $\mathcal{U}_{A\rightarrow
BE}$ satisfying the assumptions of Section~\ref{sec:channel-assumptions} and in the set $\mathcal{S}$. It is possible to estimate the Shannon mutual information $I(X;Y)$ from the measurement data of Alice and Bob, but we are left with the  following
optimization problem for Eve's Holevo information:
\begin{equation}
\sup_{\mathcal{U}_{A\rightarrow BE} \in \mathcal{S}}\chi(Y;E)_{\mathcal{E}_{\rho}},
\end{equation}
where the optimization is with respect to all collective attacks of Eve consistent with the measurement data of Alice and Bob, and the subscript notation $\mathcal{E}_{\rho}$ indicates that the Holevo information $\chi(Y;E)$ between Bob's measurement outcome
and Eve's quantum system is being computed with respect to the  following ensemble:%
\begin{equation}
\mathcal{E}_{\rho}\equiv\{r(y),\rho_{E}^{y}\},
\end{equation}
where%
\begin{align}
r(y)  & \equiv \sum_{x}r(x,y), \\
\rho_{E}^{y}  & \equiv \sum_{x}r(x|y)\rho_{E}^{x,y}\notag \\
& =\sum_{x}\frac{r(x|y)}{r(y|x)}\operatorname{Tr}_{B}\{(\Lambda_{B}^{y}\otimes
I_{E})\mathcal{U}_{A\rightarrow BE}(|\alpha_{x}\rangle\langle\alpha_{x}|_A)\}\notag \\
& =\frac{1}{r(y)}\operatorname{Tr}_{B}\{(\Lambda_{B}^{y}\otimes I_{E}%
)\mathcal{U}_{A\rightarrow BE}(\overline{\rho}_A)\}.
\end{align}
From the data processing inequality for trace distance (under the action of
the isometric channel $\mathcal{U}_{A\rightarrow BE}$ and Bob's measurement channel), we
find that%
\begin{align}
\varepsilon
& \geq\frac{1}{2}\left\Vert \overline{\rho}-\theta(N_{S})\right\Vert
_{1}\\
& \geq\frac{1}{2}\int dy\left\Vert r(y)\rho_{E}^{y}-r^G(y)\theta_{E}^{y}%
(N_{S})\right\Vert _{1},
\end{align}
where%
\begin{align}
r^G(y)  & \equiv \operatorname{Tr}\{(\Lambda_{B}^{y}\otimes I_{E})\mathcal{U}%
_{A\rightarrow BE}(\theta(N_{S}))\},\\
\theta_{E}^{y}(N_{S})  & \equiv \frac{1}{r^G(y)}\operatorname{Tr}_{B}\{(\Lambda
_{B}^{y}\otimes I_{E})\mathcal{U}_{A\rightarrow BE}(\theta(N_{S}))\}.
\end{align}
We then define the following ensemble as that which would arise had Alice employed a Gaussian modulation at the channel input:
\begin{equation}
\mathcal{E}_{\theta}=\{r^G(y),\theta_{E}^{y}\}.
\end{equation}

At this point, we invoke the fourth assumption from Section~\ref{sec:channel-assumptions}: if the mean energy of the input
state to the channel $\operatorname{Tr}_{B}\circ \mathcal{U}_{A\rightarrow BE}$
is fixed at some finite mean photon number $\kappa\in\lbrack0,\infty)$, then the mean energy of
the output state is no larger than $\kappa^{\prime}(\kappa) \in\lbrack0,\infty)$. Supposing that $H_E$ is the Hamiltonian for Eve's system $E$ satisfying the properties stated in the fourth assumption from Section~\ref{sec:channel-assumptions}, by
applying the continuity bound given in  \cite[Proposition~27]{Shirokov}, we find that%
\begin{equation}
\chi(Y;E)_{\mathcal{E}_{\rho}}\leq \chi(Y;E)_{\mathcal{E}_{\theta}}+f(\varepsilon
,P),
\end{equation}
where $P$ is an upper bound on the mean energy of the states
$\operatorname{Tr}_{B}\circ\mathcal{U}_{A\rightarrow BE}(\overline{\rho}_A)$ and
$\operatorname{Tr}_{B}\circ\mathcal{U}_{A\rightarrow BE}({\theta(N_S)})$
and $f(\varepsilon
,P)$ is a function of $\varepsilon$ and $P$, given in \cite{Shirokov}, with
the property that
\begin{equation}
    \lim_{\varepsilon\rightarrow0}f(\varepsilon,P)=0.
\end{equation}
In particular, the function $f(\varepsilon,P)$ is given by
\begin{multline} \label{eq:continuity_error}
f(\varepsilon,P) \equiv 
\varepsilon\left(2t+r_{\varepsilon}(t)\right)S(\theta_E(P/\varepsilon t))\\
+2g(\varepsilon r_\varepsilon (t))+2h(\varepsilon t),  \end{multline}
    for any $t\in (0,\frac{1}{2\varepsilon}]$, where
\begin{align}
    r_\varepsilon(t)& \equiv (1+t/2)/(1-\varepsilon t),
    \\
    g(x)& \equiv (x+1)\log_2(x+1)-x\log_2(x), \label{eq:g-func} \\
    h(x) & \equiv
    -x\log_2 (x) - (1-x)\log_2(1-x),
\end{align}
and $S(\theta_E(P/\varepsilon t))$ is the entropy of a thermal state $\theta_E(P/\varepsilon t)$ of Eve's system with mean energy $P/\varepsilon t$.
Due to this uniform bound, we can then  apply suprema to find
that%
\begin{equation}\label{eqn:Holevo-new-bound}
\sup_{\mathcal{U}_{A\rightarrow BE} \in \mathcal{S}}\chi(Y;E)_{\mathcal{E}_{\rho}}\leq \\
\sup_{\mathcal{U}_{A\rightarrow BE} \in \mathcal{S}}\chi(Y;E)_{\mathcal{E}_{\theta}%
}+f(\varepsilon,P),
\end{equation}
with the optimizations again taken with respect to collective attacks of Eve consistent with the measurement data of Alice and Bob. The lower bound on the key rate is then given as 
\begin{equation}
    K \geq I(X;Y)-\sup_{\mathcal{U}_{A\rightarrow BE} \in \mathcal{S}}\chi(Y;E)_{\mathcal{E}_{\theta}%
}-f(\varepsilon,P).
    \label{eq:key_rate}
\end{equation}
The Shannon mutual information between $X$ and $Y$, i.e., the term $I(X;Y)$, can be calculated from the observed data, as mentioned previously. The term $f(\varepsilon,P)$, introduced due to the continuity of Holevo information, can be estimated from \eqref{eq:continuity_error}. Obtaining an upper bound on the remaining term, the Holevo information
$\sup_{\mathcal{U}_{A\rightarrow BE} \in \mathcal{S}} \chi(Y;E)_{\mathcal{E}_{\theta}}$,
still requires further development, which we detail in the next section. 

\section{Channel estimation}

\label{sec:channel_estimation}

The main objective of this section is to obtain an upper bound on the remaining term, the Holevo information $\sup_{\mathcal{U}_{A\rightarrow BE} \in \mathcal{S}}\chi(Y;E)_{\mathcal{E}_{\theta}}$. The approach that we take to obtain an upper bound can be divided into three parts: estimation of parameters from the actual protocol described in Section~\ref{sec:protocol}, using these to bound the parameters that would result if a Gaussian-modulation protocol had been employed instead, and finally using these last estimates to bound the Holevo information $\sup_{\mathcal{U}_{A\rightarrow BE} \in \mathcal{S}}\chi(Y;E)_{\mathcal{E}_{\theta}}$ from above. 

\subsection{Estimation of parameters from the actual discrete-modulation protocol}

Alice and Bob calculate the parameters $\gamma_{11}$, $\gamma_{12}$, and $\gamma_{22}$ given in \eqref{eq:gamma11-param}--\eqref{eq:gamma22-param} or in \eqref{eq:gamma11-param-hetero}--\eqref{eq:gamma22-param-hetero}, depending on Bob's measurement, as described in Section~\ref{sec:protocol}.
Then the set~$\mathcal{S}$ discussed in Section~\ref{sec:proof} consists of all of the isometric channels $\mathcal{U}^{\mathcal{N}}_{A\to BE}$ that are consistent with the calculated values of $\gamma_{11}$, $\gamma_{12}$, and $\gamma_{22}$. In this way, Alice and Bob characterize the attack by Eve.


Since we are operating in the asymptotic regime, such that the number $n$ of rounds is large, it follows that the number $\delta n$ of channel estimation rounds is also large. Additionally, since Eve is employing a collective attack and the protocol has an i.i.d.~structure, it follows that the parameters $\gamma_{11}$, $\gamma_{12}$, and $\gamma_{22}$ are given exactly as the expectation of particular random variables.

To determine these random variables, we now give  exact expressions for the constellation $\{\alpha_x\}_{x=1}^{m^2}$ and distribution $r_X(x)$ that are used in the protocol. We begin by recalling the Gauss-Hermite approximation to the normal distribution with zero mean and unit variance, which reproduces the first $2m-1$ moments of the Gaussian distribution \cite[Section~3.6]{StoerBulirsch2002}. Let $H_m$ be the $m$th Hermite polynomial, and let $L_m$ be a random variable with $m$ realizations $l_{wm}$, with  probability distribution given by $r_{L_m}(l_{wm})$, where $w\in\{1,2,\ldots, m\}$. Then, as defined in \cite{Verdu}, the values $l_{wm}$ are set to the roots of the Hermite polynomial $H_m$, and the probability distribution $r_{L_m}(l_{wm})$ is defined as
\begin{equation}\label{eqn:gauss_harmite}
    r_{L_m}(l_{wm})\equiv \frac{(m-1)!}{m H^2_{m-1}(l_{wm})}.
\end{equation}

\begin{figure*}
    \centering
    \subfigure[]{\includegraphics[width=0.3\textwidth]{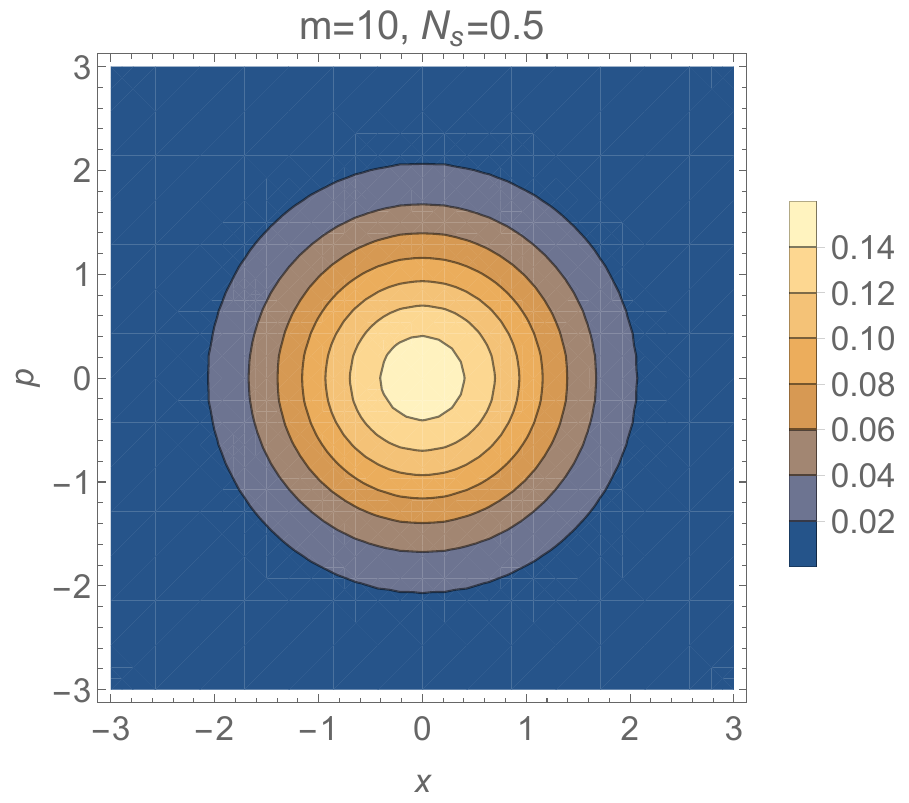}} 
    \subfigure[]{\includegraphics[width=0.3\textwidth]{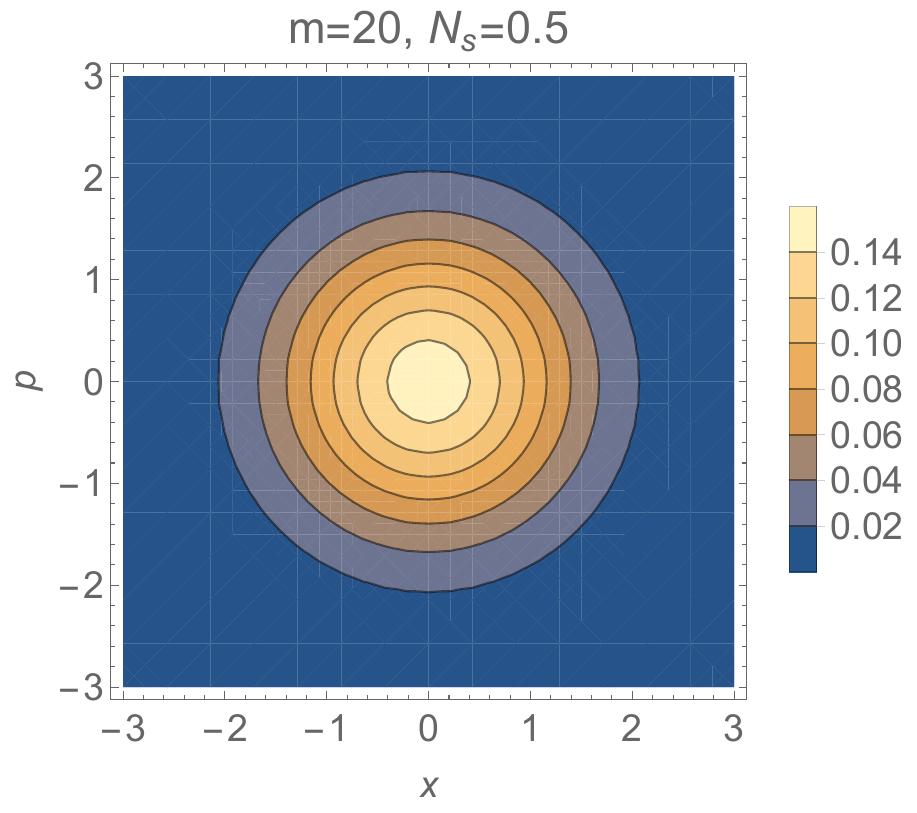}} \\
    \subfigure[]{\includegraphics[width=0.3\textwidth]{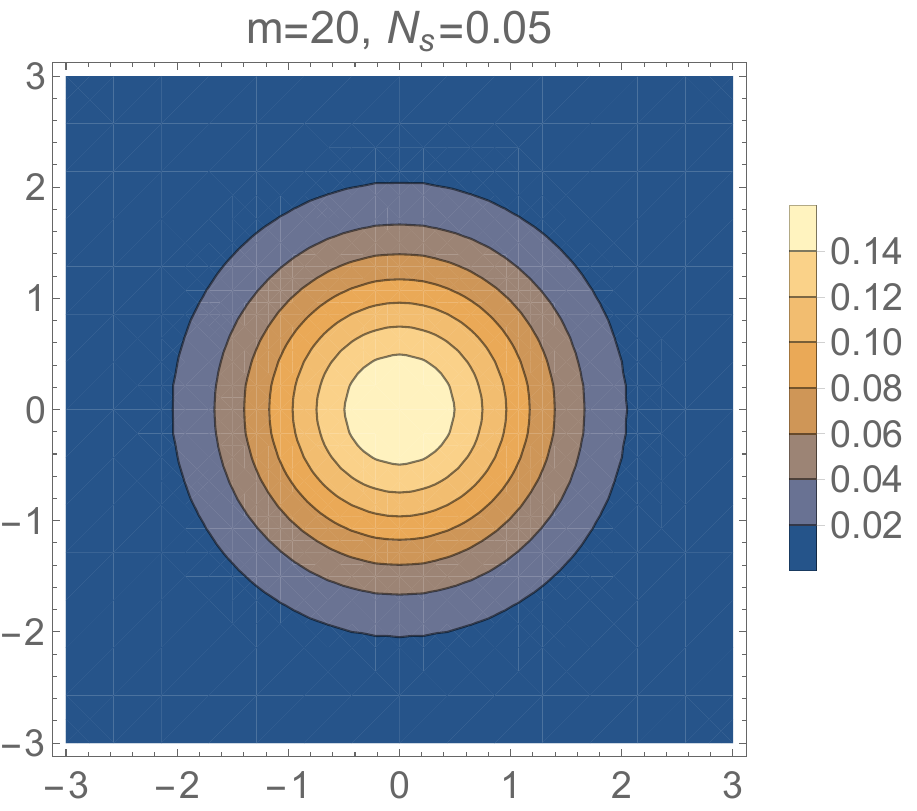}}
    \subfigure[]{\includegraphics[width=0.3\textwidth]{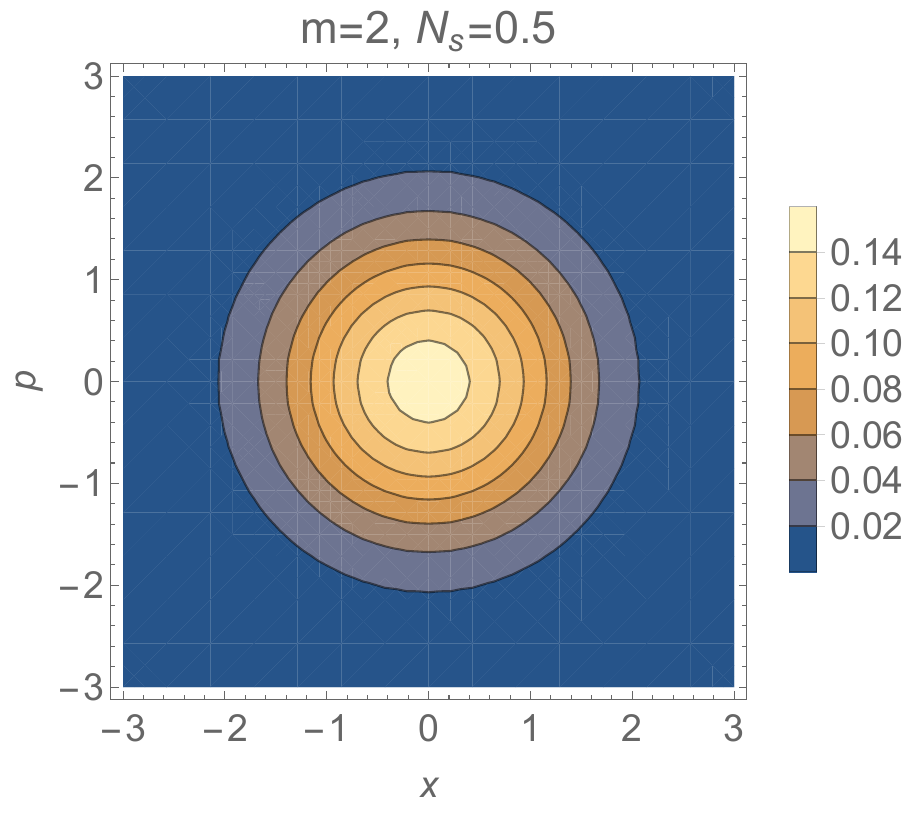}}
    \caption{ In this figure, we plot the phase space distribution for the discrete modulation state $\bar{\rho}$.  (a) $M=10, N_s= .5$ (b) $M=20, N_s= .5$ (c) $M=20, N_s= .05$ (d) $M=2, N_s= .5$.}
    \label{fig:discrete-mod-figure}
\end{figure*}

The $P$-function of a thermal state with mean photon number $N_S$ is a circularly symmetric complex Gaussian \cite{GK04}. Following \cite{Renes}, we approximate the real and imaginary parts of the thermal-state $P$-function individually by the constellation described above. Specifically, we
choose $q_{wm}$ for $w\in \{1,\ldots, m\}$ such that  the sequence $\{q_{wm} / \sqrt{N_S}\}_w$ is equal to the zeros of the Hermite polynomial $H_m$, and we choose $p_{tm}$ for $t\in \{1,\ldots, m\}$ such that the sequence $\{p_{tm}/ \sqrt{N_S}\}_{t}$ is equal to the zeros of the Hermite polynomial $H_m$. Then the constellation is given by the following distribution:
\begin{align}
    r_X(x) & =r_{X}(\alpha_{x})\\
    & = r_{X}\!\left(\frac{q_{wm}+ip_{tm}}{\sqrt{2}}\right)  \\
    & = r_{L_m}\!\left(\frac{q_{wm}}{\sqrt{N_S}}\right)\ 
    r_{L_m}\!\left(\frac{p_{tm}}{\sqrt{N_S}} \right) \\
    & \equiv r_{Q_A}(q_{wm})\ 
    r_{P_A}(p_{tm}),
\end{align}
where $x = (w,t) \in\{1,\ldots, m\}\times \{1,\ldots, m\}$. The factor $\sqrt{N_S}$ is a scaling factor incorporated so that the mean photon number of the expected density operator for the resulting constellation  is equal to the mean photon number of the thermal state $\theta(N_S)$. The phase space distribution for several discrete modulated states is give in Figure~\ref{fig:discrete-mod-figure}.

Let $Q_A$ denote the discrete random variable with realizations $q_A \in \mathbb{R}$, taking values $q_{wm}$ and having a probability distribution as detailed above. Let $Q_B$ denote the random variable associated to Bob's homodyne measurement outcome of the position-quadrature operator, taking values in $\mathbb{R}$. Then, for characterizing the isometric channels $\mathcal{U}_{A\rightarrow BE}$ in $\mathcal{S}$, Alice and Bob calculate the parameters
$\gamma_{11}$, $\gamma_{12}$, and $\gamma_{22}$ from their data. Due to the fact that we are operating in the asymptotic regime (with no finite-size statistical effects), the following equalities hold for protocols with homodyne detection
\begin{align}
\gamma_{11} & =\mathbb{E}\left[(Q_A - \mathbb{E}[Q_A])^2\right], \label{eq:DM-gam-11} \\
\gamma_{12} & =\mathbb{E}\left[(Q_A -  \mathbb{E}[Q_A])(Q_B -  \mathbb{E}[Q_B])\right],\label{eqn:covarinace-paramter}\\
\gamma_{22} & =\mathbb{E}\left[(Q_B- \mathbb{E}[Q_B])^2\right].\label{eq:DM-gam-22}
\end{align}

Now consider the discrete-modulation protocols with heterodyne detection. Let $Q_A$ denote the discrete random variable with realizations $q_A \in \mathbb{R}$, taking values $q_{wm}$ and having a probability distribution as detailed above. Let $P_A$ denote the discrete random variable with realizations $p_A \in \mathbb{R}$, taking values $p_{wm}$ and having a probability distribution as detailed above. Let $Q_B$ denote the random variable associated to Bob's heterodyne measurement outcome of the position-quadrature operator, taking values in $\mathbb{R}$. Let $P_B$ denote the random variable associated to Bob's heterodyne measurement outcome of the momentum-quadrature operator, taking values in $\mathbb{R}$. Then, in the asymptotic regime the following equalities hold for protocols with heterodyne measurement
\begin{align}
\gamma_{11} & =\mathbb{E}\left[(Q_A - \mathbb{E}[Q_A])^2\right] = \mathbb{E}\left[(P_A - \mathbb{E}[P_A])^2\right], \label{eq:DM-gam-11-hetero} \\
\gamma_{12} & =\frac{1}{2}\Big(\mathbb{E}\left[(Q_A -  \mathbb{E}[Q_A])(Q_B -  \mathbb{E}[Q_B])\right]+\nonumber\\& \qquad \qquad \mathbb{E}\left[(P_A -  \mathbb{E}[P_A])(P_B -  \mathbb{E}[P_B])\right]\Big),\label{eqn:covarinace-paramter-hetero}\\
\gamma_{22} & =\frac{1}{2}\left(\mathbb{E}\left[(Q_B- \mathbb{E}[Q_B])^2\right]+\mathbb{E}\left[(P_B- \mathbb{E}[P_B])^2\right]\right).\label{eq:DM-gam-22-hetero}
\end{align}
Due to the symmetry of the protocol, \eqref{eqn:covarinace-paramter-hetero} and \eqref{eq:DM-gam-22-hetero} can be simplified as
\begin{align}
    \gamma_{12}&=\mathbb{E}\left[(Q_A -  \mathbb{E}[Q_A])(Q_B -  \mathbb{E}[Q_B])\right],\\
    \gamma_{22}&=\mathbb{E}\left[(Q_B- \mathbb{E}[Q_B])^2\right].
\end{align}
As stated previously, Alice estimates $\gamma_{11}$ from her preparation data, while Bob estimates $\gamma_{22}$ from his measurement data. Alice calculates $\gamma_{12}$ from the data that is publicly published by Bob. Then $\mathcal{S}$ is the set of isometric channels $\mathcal{U}_{A\rightarrow BE}$ that fulfill the constraints in Section~\ref{sec:channel-assumptions} and produce the observed values of $\gamma_{12}$ and $\gamma_{22}$. As a consequence, Alice and Bob deduce that the attack by Eve yields the observed values of $\gamma_{12}$ and $\gamma_{22}$. In this way, they are able to restrict the possible attacks that could have been performed by Eve.

\subsection{Estimation of parameters for a hypothetical Gaussian-modulation protocol}
\label{sec:est-parameters-hypothetical}

Now notice that the remaining Holevo information
$\sup_{\mathcal{U}_{A\rightarrow BE} \in \mathcal{S}}\chi(Y;E)_{\mathcal{E}_{\theta}}$ from \eqref{eq:key_rate} 
that we want to bound from above is calculated for a thermal state $\theta(N_S)$ sent over an isometric channel $\mathcal{U}_{A\rightarrow BE}$ in the set $\mathcal{S}$ and Bob performing homodyne or heterodyne detection. Therefore, we want to obtain an estimate of the parameters
    $\gamma_{11}^G$,   $\gamma_{12}^G$, and $\gamma_{22}^G$, which are defined analogously to 
    \eqref{eq:DM-gam-11}--\eqref{eq:DM-gam-22}, but with the initial random variable $Q_A$ replaced by a Gaussian random variable with mean zero and variance equal to $N_S$.
The parameters $\gamma_{11}^G$,   $\gamma_{12}^G$, and $\gamma_{22}^G$ are those that would be observed in a Gaussian modulation protocol when the average channel input of Alice is a thermal state $\theta(N_S)$ instead of $\overline{\rho}$. A hypothetical Gaussian-modulation protocol refers to a protocol in which the average state that Alice sends is a thermal state $\theta(N_S)$ instead of $\overline{\rho}$. This protocol is not carried out by Alice and Bob experimentally, but the parameters $\gamma_{11}^G,\gamma_{12}^G,\gamma_{22}^G$ corresponding to the hypothetical Gaussian modulation protocol are inferred from the discrete-modulation protocol. 

In order to bound the values of the parameters that would be obtained in a Gaussian-modulation protocol with Eve's attack taken from the set $\mathcal{S}$, we can employ the parameters that are observed in the discrete-modulation protocol. Before we do so, let us recall the definition of the $\chi^2$ divergence of two states $\rho$ and $\sigma$ as $\chi^2(\rho,\sigma)\equiv \Tr[(\rho \sigma^{-1/2})^2]-1$ \cite{Temme}. Then we have the following proposition: 
\begin{proposition}\label{prop:estimategaussian}
Let $\overline{\rho}= \sum_x r_X(x) \op{\alpha_x}$, where
\begin{align}
\alpha_x & =\frac{q_A+ip_A}{\sqrt{2}}, \\
r_X(x) & = r_{Q_A}(q_A)\, r_{P_A}(p_A), \\
\theta_{N_S} & = \int dx\, r^G_X(x)\op{\alpha_x},
\end{align}
and $r^G(x)$ is the $P$-function of a thermal state with mean photon number $N_S$. If  $\sqrt{\chi^2(\overline{\rho},\theta(N_S))} \leq \varepsilon^2$, and Eve's attacks fulfill the constraints in Section~\ref{sec:channel-assumptions},
then
\begin{align}
\gamma_{11} &= \gamma_{11}^G, \label{app:eq:11-equality}\\
|\gamma_{22}-\gamma_{22}^G| &\leq \varepsilon_1,\label{app:eq:12-inequality}\\
|\gamma_{12}-\gamma_{12}^G| &\leq \varepsilon_2,\label{app:eq13-inequality}
\end{align}
where 
\begin{align}
       \varepsilon_1 & \equiv \varepsilon\cdot\left(1+c_1\right)\cdot\sqrt{\mathbb{E}\left[(Q_B- \mathbb{E}[Q_B])^4\right]},
\end{align}
for some constant $c_1>0$ and
\begin{multline}
\varepsilon_2 \equiv \sum_{k=0}^{2m-2}\sum_{l=2m}^{K}\mu_{kl}\left|\eta^G(q_A,k+1)\left(\eta^G(p_A,l)-\eta(p_A,l)\right)\right|
\\
+ \sum_{k=2m-1}^{K}\sum_{l=0}^{2m-1}\mu_{kl}\left|\eta^G(p_A,l)\left(\eta^G(q_A,k+1)-\eta(q_A,k+1)\right)\right|
\\
+\sum_{k=2m-1,l=2m}^{K}\mu_{kl}\left|\eta^G(p_A,l)\eta^G(q_A,k+1)-\right.
\\
\left.\eta(p_A,l)\eta(q_A,k+1)\right|,
\end{multline}
where $\mu_{kl}$ is an arbitrary function for $k\leq 2m-2,\ l\leq2m-1$ and is equal to $\exp(-a(k+l))$ otherwise. We also define the following quantities
\begin{align}
    \eta^G(q_A,k) & \equiv \mathbb{E}_{r^G_{Q_A}}[Q_A^{k}],\\
    \eta(q_A,k) & \equiv \mathbb{E}_{r_{Q_A}}[Q_A^{k}],\\
    \eta^G(p_A,k) & \equiv \mathbb{E}_{r^G_{P_A}}[P_A^{k}],\\
    \eta(p_A,k) & \equiv \mathbb{E}_{r_{P_A}}[P_A^{k}].
\end{align}

\end{proposition}

Our proof of \eqref{app:eq:12-inequality} relies mainly on the properties of trace distance, invoking the Cauchy--Schwarz inequality and the assumption that the fourth moment of the channel output is bounded. Our proof of \eqref{app:eq13-inequality} relies mainly on the properties of Gauss-Hermite distribution, and on the last assumption in Section~\ref{sec:channel-assumptions}.  For details, please refer to Appendix~\ref{app:estimategaussian}. By invoking Proposition~\ref{prop:estimategaussian}, we conclude that $\gamma^G_{22}\in \left[\gamma_{22}-\varepsilon_1, \gamma_{22}+\varepsilon_1\right]$, and $\gamma^G_{12}\in \left[\gamma_{12}-\varepsilon_2, \gamma_{12}+\varepsilon_2\right]$, where $\varepsilon_{1}$ and $\varepsilon_{2}$ are defined above. 

Now, consider the following scenario corresponding to an entanglement-based (EB) QKD protocol: Alice prepares a two-mode squeezed vacuum state
$\psi(\bar{n})_{RA} = |\psi(\bar{n})\rangle \langle \psi(\bar{n})|_{RA}$ where
\begin{equation}
    |\psi(\bar{n})\rangle_{RA} \equiv
    \frac{1}{\sqrt{\bar{n}+1}}
    \sum_{n=0}^{\infty} \sqrt{\left(\frac{\bar{n}}{\bar{n}+1}\right)^n} \ket{n}_R \ket{n}_A, \label{eq:TMSV}
\end{equation}
with $\bar{n}\geq 0$. She applies a  phase $e^{-i \hat{n} \pi k/2}$ to her channel input mode $A$, with $k \in \{0,1,2,3\}$ selected uniformly at random, and she sends the system $A$  to Bob over an isometric channel $\mathcal{U}^{\mathcal{N}}_{A \to BE}$ selected from the set $\mathcal{S}$. She also communicates $k$ to Bob over an authenticated public classical channel. Bob then applies the inverse phase $e^{-i \hat{n} \pi k/2}$. Both Alice and Bob then discard the value of~$k$. Let
$\rho_{RB}$ denote
the state shared by Alice and Bob at the end, so that the reduced channel $\mathcal{N}_{A\to B}$ has been phase symmetrized due to the protocol above and with $\overline{\mathcal{N}}_{A\to B}$ defined as  in \eqref{eq:phase-sym-channel}:
\begin{equation}
    \rho_{RB} \equiv \overline{\mathcal{N}}_{A\to B}(\psi(\bar{n})_{RA}). \label{eq:rho-EB-protocol}
\end{equation}
Due to the symmetries of the two-mode squeezed vacuum state $\psi(\bar{n})_{RA}$ as well as those of the phase-symmetrized channel $\overline{\mathcal{N}}_{A\to B}$, it follows that the covariance matrix of the state $\rho_{RB}$ has the following form:
\begin{equation}
    \begin{bmatrix}
\gamma_{11}^{\operatorname{EB}} \, \mathbb{I}_2 & \gamma_{12}^{\operatorname{EB}} \, R(\theta)\\
\gamma_{12}^{\operatorname{EB}}\, R(\theta) & \gamma_{22}^{\operatorname{EB}}\,  \mathbb{I}_2
\end{bmatrix},
\end{equation}
where $\gamma_{11}^{\operatorname{EB}}, \gamma_{12}^{\operatorname{EB}}, \gamma_{22}^{\operatorname{EB}} \in \mathbb{R}$ such that the matrix above is a legitimate quantum covariance matrix \cite{S17}, the matrix $\mathbb{I}_2$ is the $2\times 2$ identity matrix, and
\begin{equation}
    R(\theta) \equiv \begin{bmatrix}
\cos(\theta) & \sin(\theta)\\
\sin(\theta) & -\cos(\theta)
\end{bmatrix},
\end{equation}
is a rotation matrix.
See Appendix~\ref{appendix_sym} for a proof of this claim.
In what follows, we assume that $\theta = 0$, due to the fact that doing so simplifies the protocol, as well as reduces the number of parameters that need to be estimated, and it furthermore does not lead to an increase in Eve's Holevo information, as discussed in \cite{L15}.
Alice then performs a heterodyne measurement on mode $R$ and Bob performs a homodyne or a heterodyne measurement on mode $B$. As mentioned above, this is the entanglement-based (EB) version of the Gaussian-modulated prepare-measure (PM) protocol with the attacks by Eve constrained to the set~$\mathcal{S}$.

Now, we want to deduce the parameters
$\gamma_{11}^{\operatorname{EB}}, \gamma_{12}^{\operatorname{EB}}, \gamma_{22}^{\operatorname{EB}}$
observed in the EB protocol from the parameters
$\gamma_{11}^{G}, \gamma_{12}^{G}, \gamma_{22}^{G}$
observed in the PM version of the Gaussian modulation protocol. As is common in the CV-QKD literature, we consider the EB protocol because it is helpful in analyzing the Holevo information $\chi(Y;E)$ that results in the prepare-measure (PM) protocol. The ``PM to EB'' mapping of the parameters is well known in the literature \cite{Laudenbach} and is given as follows for protocol where Bob performs homodyne detection:
\begin{align}
\gamma_{11}^{\operatorname{EB}}&=\gamma_{11}^{G}+1=\gamma_{11}+1, \\
\gamma_{22}^{\operatorname{EB}}&=\gamma_{22}^{G}\in \left[\gamma_{22}-\varepsilon_1, \gamma_{22}+\varepsilon_1\right], \\
\gamma_{12}^{\operatorname{EB}}&=\sqrt{\frac{\gamma_{11}+2}{\gamma_{11}}}\gamma_{12}^{G}\notag \\
& \in\left[\sqrt{\frac{\gamma_{11}+2}{\gamma_{11}}}(\gamma_{12}-\varepsilon_2), \sqrt{\frac{\gamma_{11}+2}{\gamma_{11}}}(\gamma_{12}+\varepsilon_2)\right].
\end{align}
For protocols where Bob performs heterodyne detection the ``PM to EB" mapping of the parameters is given by
\begin{align}
\gamma_{11}^{\operatorname{EB}}&=\gamma_{11}^{G}+1=\gamma_{11}+1, \\
\gamma_{22}^{\operatorname{EB}}&=2\gamma_{22}^{G}-1\in \left[2\gamma_{22}-1-\varepsilon_1, 2\gamma_{22}-1+\varepsilon_1\right], \\
\gamma_{12}^{\operatorname{EB}}&=\sqrt{\frac{2(\gamma_{11}+2)}{\gamma_{11}}}\gamma_{12}^{G}\notag \\
& \in\left[\sqrt{\frac{2(\gamma_{11}+2)}{\gamma_{11}}}(\gamma_{12}-\varepsilon_2), \sqrt{\frac{2(\gamma_{11}+2)}{(\gamma_{11}}}(\gamma_{12}+\varepsilon_2)\right].
\end{align}

Let $\Sigma$ denote the set of quantum states 
$\rho_{RB}$
that have covariance matrix of the following form:
\begin{equation}
\begin{bmatrix}
\gamma_{11}^{\operatorname{EB}} \, \mathbb{I}_2 & \gamma_{12}^{\operatorname{EB}}\sigma_Z\\
\gamma_{12}^{\operatorname{EB}}\sigma_Z & \gamma_{22}^{\operatorname{EB}}\,  \mathbb{I}_2
\end{bmatrix}.
\label{eq:EB-cov-matrix}
\end{equation}

\subsection{Upper bound on Eve's Holevo information}

By applying purification techniques of quantum information theory, the following equality holds
\begin{equation}\label{eqn:purifying-holevo}
\chi(Y;E)_{\mathcal{E}_\theta} = H(RB)_{\rho} - H(R|Y)_{\{p(y),\rho^y\}_y},    
\end{equation}
for $\rho_{RB}$ the state in \eqref{eq:rho-EB-protocol} and  $\{p(y),\rho_R^y\}_y$ the ensemble resulting from Bob performing a position-quadrature homodyne detection, or a heterodyne detection on the state $\rho_{RB}$.
As a consequence, 
the task of obtaining an upper bound on $\sup_{\mathcal{U}_{A\rightarrow BE} \in \mathcal{S}}\chi(Y;E)_{\mathcal{E}_{\theta}}$ can be accomplished by obtaining an upper bound on $\sup_{\rho_{RB}\in \Sigma}\left(H(RB)_{\rho}-H(R|Y)_{\{p(y),\rho^y\}_y}\right)$.

We then invoke the extremality of Gaussian states \cite{Wolf,Patron2006}, from which we infer that the Holevo information is optimized by a Gaussian state $\rho_{RB}^G$ having the same covariance matrix as $\rho_{RB}$. Therefore, we obtain the following: 
\begin{multline}
 \sup_{\rho_{RB}\in \Sigma}\left(H(RB)_{\rho}-H(R|Y)_{\{p(y),\rho^y\}}\right)\\
 = \sup_{\rho^G_{RB}\in \Sigma}\left(H(RB)_{\rho^G}-H(R|Y)_{\{p^G(y),\rho^{y,G}\}}\right),
\end{multline}
where $\{p^G(y),\rho_R^{y,G}\}$ is the ensemble obtained if Bob performs a homodyne/heterodyne measurement on mode $B$ of $\rho_{RB}^G$.

Then, Eve's Holevo information can be calculated as follows:
\begin{multline}\label{eq:eve-holevo-information}
H(RB)_{\rho^G}-H(R|Y)_{\{p^G(y),\rho^{y,G}\}}\\
     = g(\nu_1)+g(\nu_2)-g(\nu_3),
\end{multline}
where the function $g(\cdot)$ is defined in \eqref{eq:g-func}, $\nu_1$ and $\nu_2$ are the symplectic eigenvalues of the covariance matrix in \eqref{eq:EB-cov-matrix}. For protocols where Bob performs homodyne detection, $\nu_3= \gamma_{11}^{\operatorname{EB}}\left(\gamma_{11}-\frac{\left(\gamma_{12}^{\operatorname{EB}}\right)^2}{\gamma_{22}^{\operatorname{EB}}+1}\right)$. For protocols where Bob performs heterodyne detection, $\nu_3=\gamma_{11}^{\operatorname{EB}}-\frac{(\gamma_{22}^{\operatorname{EB}})^2}{\gamma_{12}^{\operatorname{EB}}+1}$.   Numerical checks, similar to those performed and stated in \cite{L15}, reveal that the Holevo information is a monotonically decreasing function of $\gamma_{12}^{\operatorname{EB}}$, and a monotonically increasing function of $\gamma_{11}^{\operatorname{EB}}$
and $\gamma_{22}^{\operatorname{EB}}$. Intuitively, the correlations between Alice and Bob are quantified by $\gamma_{12}^{\operatorname{EB}}$, so that increasing this parameter decreases Eve's Holevo information. 

Therefore, we conclude that the Holevo information for protocols where Bob performs homodyne detection is no larger than that achieved by a Gaussian state $\rho_{RB}$ that has a covariance matrix as follows:
\begin{equation} \label{eq:covariance_matrix_eb}
\begin{bmatrix}
(\gamma_{11}+1)\mathbb{I} & \sqrt{\frac{\gamma_{11}+2}{\gamma_{11}}}(\gamma_{12}-\varepsilon_2)\sigma_Z\\
\sqrt{\frac{\gamma_{11}+2}{\gamma_{11}}}(\gamma_{12}-\varepsilon_2)\sigma_Z & (\gamma_{22}+\varepsilon_1) \mathbb{I}%
\end{bmatrix}.
\end{equation}

The Holevo information for protocols where Bob performs heterodyne detection is no larger than that achieved by a Gaussian state $\rho_{RB}$ that has a covariance matrix as follows:
\begin{equation} \label{eq:covariance_matrix_eb1}
\begin{bmatrix}
(\gamma_{11}+1)\mathbb{I} & \sqrt{\frac{2(\gamma_{11}+2)}{\gamma_{11}}}(\gamma_{12}-\varepsilon_2)\sigma_Z\\
\sqrt{\frac{2(\gamma_{11}+2)}{\gamma_{11}}}(\gamma_{12}-\varepsilon_2)\sigma_Z & (2\gamma_{22}-1+\varepsilon_1) \mathbb{I}%
\end{bmatrix}.
\end{equation}
With this, we conclude our goal of obtaining an upper bound on the remaining term $\sup_{\mathcal{U}_{A\rightarrow BE} \in \mathcal{S}}\chi(Y;E)_{\mathcal{E}_{\theta}}$. 

In Appendix~\ref{sec:appendix-alternative}, we give an alternative method to upper bound the Holevo information $\sup_{\mathcal{U}_{A\rightarrow BE} \in \mathcal{S}}\chi(Y;E)_{\mathcal{E}_{\rho}}$ in \eqref{eqn:Holevo-new-bound}. The proposed method does not depend on the parameters $c_1$, $K$ and $a$; however, it seems to be numerically intensive.

\section{Numerical results for a lossy thermal bosonic channel}

\label{sec:numerics}

\begin{figure*}
    \centering
    \includegraphics[width=.7\linewidth]{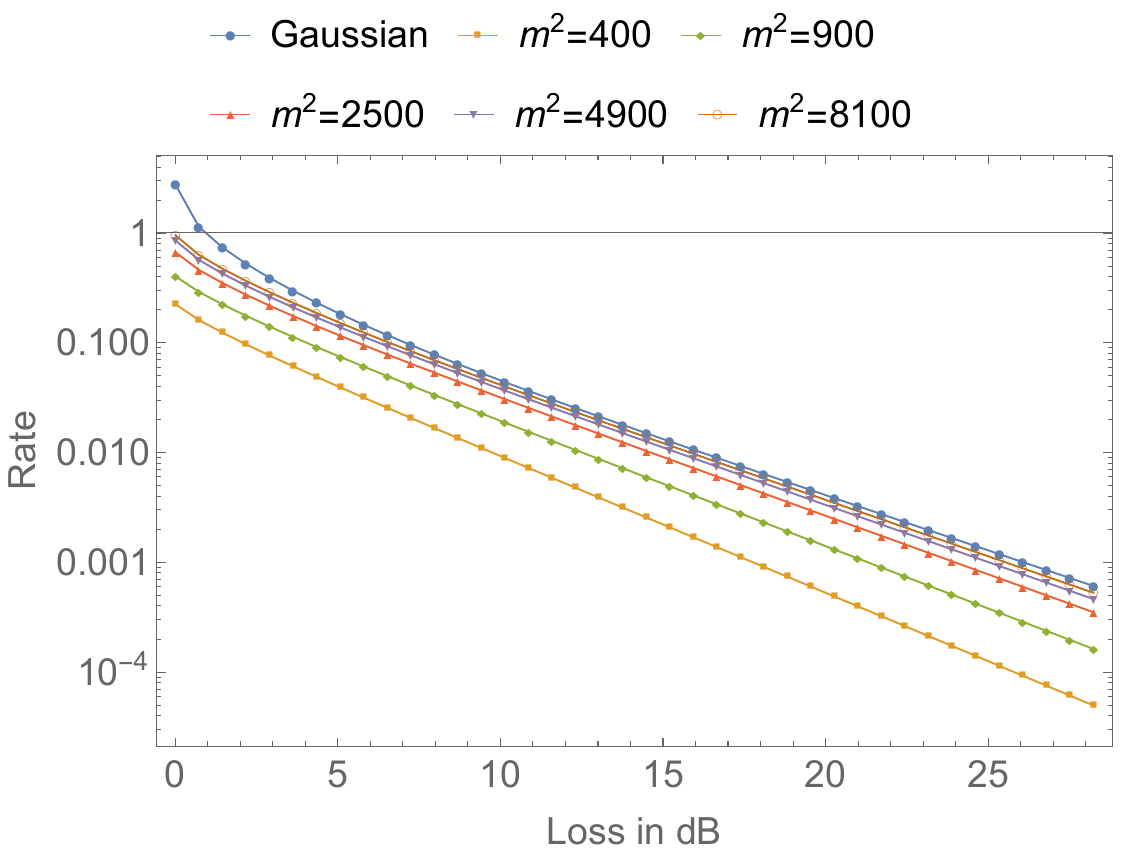}
    \caption{In this figure, we plot the lower bounds on the key rates for various constellation size $m^2$ of the discrete-modulation protocol considered in Section~\ref{sec:protocol} with the underlying channel as a lossy bosonic channel} 
    \label{fig:plots1}
\end{figure*}
\begin{figure*}
    \centering
    \includegraphics[width=1\linewidth]{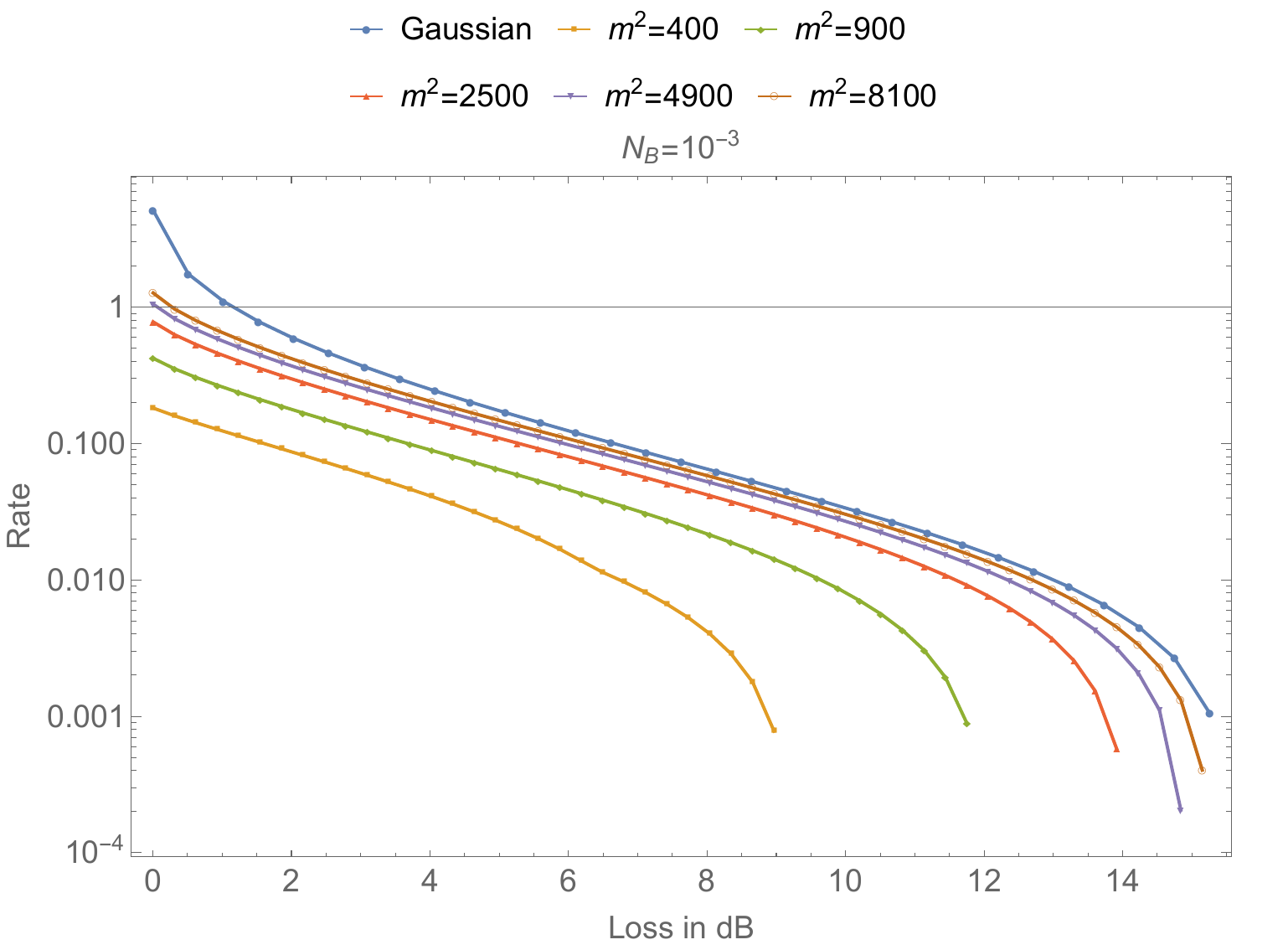}
    \caption{In this figure, we plot the lower bounds on the key rate for various constellation size $m^2$ of the discrete-modulation protocol considered in Section~\ref{sec:protocol} with the underlying channel as a lossy thermal bosonic channel with thermal noise $N_B =10^{-3}$.}
    \label{fig:plots2}
\end{figure*}

\begin{figure*}
    \centering
    \includegraphics[width=1\linewidth]{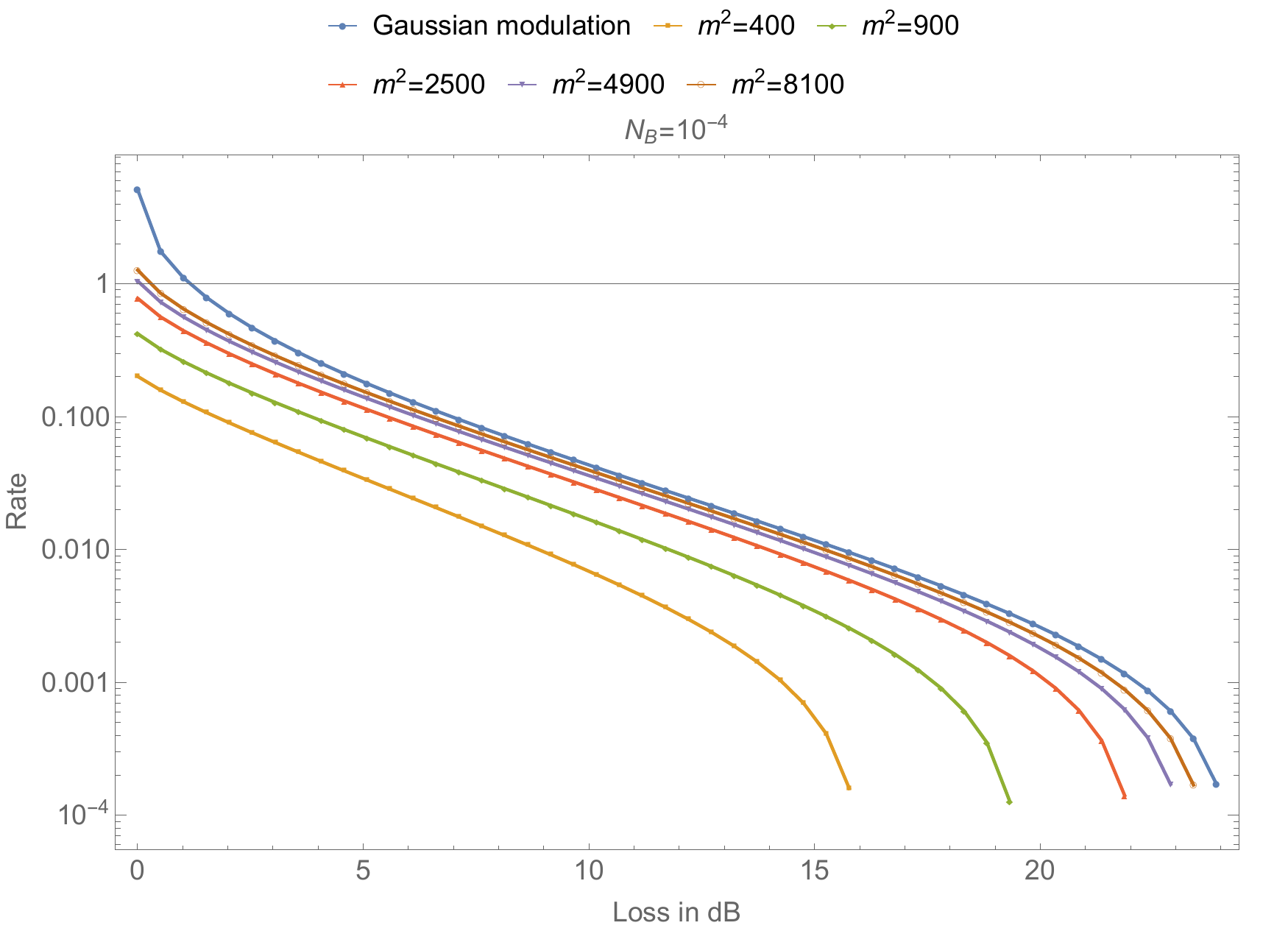}
    \caption{In this figure, we plot the lower bounds on the key rate for various constellation size $m^2$ of the discrete-modulation protocol considered in Section~\ref{sec:protocol} with the underlying channel as a lossy thermal bosonic channel $N_B =10^{-4}$.}
    \label{fig:plots3}
\end{figure*}

We now proceed with calculating the various terms in \eqref{eq:key_rate} for a Gauss-Hermite constellation of size $m^2$ and for a lossy thermal bosonic channel of transmissivity $\eta\in[0,1]$ and $N_B \geq 0$. This allows for determining the performance of the discrete-modulation CV-QKD protocol when the underlying channel is a lossy thermal channels (however, keep in mind that Alice and Bob are not aware of this when executing the protocol).

The first term that we need to calculate is the Shannon mutual information $I(X;Y)$. Here, $X$ is a random variable that encodes the choice of coherent state, and $Y$ is the random variable that is associated with the measurement result. For discrete-modulation protocols with homodyne detection and the underlying channel as the pure-loss channel, we use the following approach: The Wigner function associated with the coherent state $\ket{\alpha_x}$ subjected to a pure-loss channel with transmissivity $\eta$ is given as
\begin{equation}
    W^{\alpha_x}_{y,p} = \frac{1}{\pi}\exp{-\left|z-\sqrt{\eta}\alpha_x\right|^2},
    \end{equation}
    where $z=q_B + ip_B$, with the real part $q_B$ corresponding to the position-quadrature phase-space variable, and the imaginary part $p_B$ corresponding to the momentum-quadrature phase-space variable.  Bob performs homodyne detection with respect to the $q_B$ quadrature, which provides the raw data for key generation. Then the various probability distributions are given as 
\begin{align}
    r_X(x)&= Q_{N_S,m}(\alpha_x),\\
    r_{Y|X}(q_B|x)&= \int_{-\infty}^{\infty} dp_B \ 
    W^{\alpha_x}_{q_B,p_B} ,\\
    r_{Y}(q_B) &= \sum_{x} r_X(x)r_{Y|X}(q_B|x).
\end{align}
With this information in hand, it is easy to calculate $I(X;Y)=H(Y)_{r}-H(Y|X)_r$. 
Now, let us calculate $I(X;Y)$ for the discrete modulation protocols with heterodyne detection. Alice sends a coherent state characterized by $\ket{\alpha_x}$ through a thermal channel characterized by $\eta$ and $N_B$. After the transmission, Bob has a displaced thermal state with the mean vector $\bar{r}_{\textrm{final}}$ and covariance matrix $\sigma_{\textrm{final}}$. These can be written as
\begin{align}
    \bar{r}_{\textrm{final}}&= \sqrt{\eta}\, \bar{r}_{\textrm{coherent}}=\sqrt{\eta}\left[\sqrt{2}q_{wm},\sqrt{2}p_{tm}\right]^T,\\
    \sigma_{\textrm{final}}& = \eta\, \sigma_{\textrm{coherent}} + (1-\eta)(2N_B+1)\mathbb{I}_2\nonumber\\
    &=\eta\, \mathbb{I}_2 + (1-\eta)(2N_B+1)\mathbb{I}_2,
\end{align}
where $\bar{r}_{\textrm{coherent}}$ is the mean vector and $\sigma_{\textrm{coherent}}$ is the covariance matrix of the coherent state $\ket{\alpha_x}=\ket{\frac{q_{wm}+ip_{tm}}{\sqrt{2}}}$. Then the various probability distributions are given as 
\begin{align}
  r_{Y|X}(q_b,p_b|q_{wm},p_{tm})= \frac{\exp[-\frac{(q_b-\sqrt{2}q_{wm})^2-(p_b-\sqrt{2}p_{tm})^2}{2(1-N_B(1-\eta))}]}{\pi \sqrt{\operatorname{Det}\left[2(1-N_B(1-\eta))\mathbb{I}_2\right]}}\\
  r_Y(q_b,p_b)= \sum_{q_{wm},p_{tm}}r_X(q_{wm},p_{tm})r_{Y|X}(q_b,p_b|q_{wm},p_{tm}).
\end{align}
With this information, we can easily calculate $I(X;Y)$. Numerically, we find that $I(X;Y)$ calculated from the above method is approximately equal to the $I(X;Y)$ that we obtain from a Gaussian modulation protocol with the underlying channel as thermal channel. We invoke this approximation in the numerics. This approximation has been proven rigorously in \cite{Renes}.

The second term that we need to calculate is the Holevo information $\sup_{\mathcal{U}_{A\rightarrow BE} \in \mathcal{S}}\chi(Y;E)_{\mathcal{E}_{\theta}%
 }$ for a key-generation protocol that uses Gaussian modulation of coherent states and homodyne/heterodyne detection. To this end, we need to calculate the parameters $\gamma_{11}$, $\gamma_{22}$, and $\gamma_{12}$ for the discrete-modulation protocol in order to obtain the covariance matrix in \eqref{eq:covariance_matrix_eb} or in \eqref{eq:covariance_matrix_eb1}. These can be calculated numerically. However, note that $\theta(N_S)$ and $\overline{\rho}$ have the same covariance matrix due to the second moment of the Gauss-Hermite approximation and Gaussian distribution being the same. Since we are considering the underlying channel as a lossy thermal bosonic channel, we can calculate the parameters $\gamma_{22}$ and $\gamma_{12}$ using the analytical formulas given in Section~7 of \cite{Laudenbach}.
 From the values of $\gamma_{12}$ and $\gamma_{22}$ we now have to estimate the parameters $\gamma_{12}^G$ and $\gamma_{22}^G$. To this end, we apply Proposition~\ref{prop:estimategaussian}. When applying Proposition~\ref{prop:estimategaussian}, it is necessary to make a choice for the parameters $c_1$, $a$ and $K$. In our example considered here, we take the conservative choices  $c_1 = 100$, $K=10^4$, and $a=5$. 

Next, we have to calculate the third term, which is the error introduced in the Holevo information $\chi(Y;E)$ and denoted by $f(\varepsilon,N_S')$ in \eqref{eq:key_rate}. To this end, we first calculate the approximation error~$\varepsilon$ defined in \eqref{eq:error}. Following \cite{Renes}, we use the $\chi^2$-distance, defined as $\chi^2(\rho,\sigma) \equiv \operatorname{Tr}[(\rho \sigma ^{-1/2})^2]-1$, and we employ the  bound $\|\rho-\sigma\|_1^2\leq\chi^2(\rho,\sigma)$, which follows from Lemma~5 of \cite{Temme} with $k=1/2$.

Let us denote an additive white Gaussian noise channel with signal to noise ratio $s$ by $W_s$. The action of $W_s$ is defined as $W_s(Z)=\sqrt{s}Z+G$, where $G$ is a normally distributed random variable with unit variance. Then, for $Z\sim \mathcal{N}(0,1)$ with distribution $P_Z$, a random variable $Z'_m$  with distribution $P_{Z_m}$ as given in \eqref{eqn:gauss_harmite}, $Y=W_s(Z)$, and $Y'_m=W_s(Z'_m)$, the $\chi^2$ distance is given as 
\begin{align}
\label{eqn:chi_classical}
    \chi^2\left(P_{Y'_m}, P_Y\right) & =2\kappa^2\sum_{k\geq m}\left(\frac{s}{1+s}\right)^{2k}\\
    & = 2\kappa^2 \frac{ (1+s)^{2}}{1+2s} \left(\frac{s}{1+s}\right)^{2m}, 
\end{align}
with $2\kappa^2 \approx 2.36$ \cite{Verdu}.

Let
\begin{align}
\overline{\rho}_m & =\sum_{x=1}^{m^2} Q_{N_S,m}(\alpha_x)\op{\alpha_x},
\end{align}
and $\theta_{N_S}$ be a thermal state of mean photon number $N_S$. Then 
\begin{align}
\label{eqn:chi_quantum}
    \chi^2(\overline{\rho}_m,\theta_{N_S}) & = \left(1+\chi^2(P_{Y_m}, P_Y)\right)^2-1,\\
    & = (1+\tau)^2-1\\
    & = \tau(2+\tau),
\end{align}
with $s= N_S/\left(\sqrt{N_S(N_S+1)}-N_S\right)$ \cite{Renes} and
\begin{equation}
    \tau \equiv 2\kappa^2\left(1+N_S\right)\left(\frac{N_S}{\sqrt{N_S(1+N_S)}}
    \right)
^{2m}.
\end{equation}
Combining   \eqref{eqn:chi_classical}
and \eqref{eqn:chi_quantum}, we obtain the following expression for the approximation error: 
\begin{equation}
\label{eq:error1}
\frac{1}{2} \|\overline{\rho}_m-\theta_{N_S}\|_1\leq     \varepsilon 
= \frac{1}{2} \sqrt{\tau (2+ \tau)}.
\end{equation}

We can then invoke  \cite[Proposition~27]{Shirokov}, which utilizes some techniques from \cite{Winter}, to obtain
\begin{multline}
\label{eqn:shirokovbound}
    f(\varepsilon,N_S)= \varepsilon\left(2t+r_{\varepsilon}(t)\right)g(P/\varepsilon t)+2g(\varepsilon r_\varepsilon (t))+2h(\varepsilon t),
\end{multline}
for any $t\in (0,\frac{1}{2\varepsilon}]$, where
\begin{align}
    r_\varepsilon(t)& =(1+t/2)/(1-\varepsilon t),
    \\
    P& =10^{7}, \\
    g(N)& =(N+1)\log_2(N+1)-N\log_2(N), \\
    h(x) & =-x\log_2(x)-(1-x)\log_2(1-x).
\end{align}

In the above, we have set $P=10^{7}$, which is an extremely conservative choice to employ with respect to the fourth assumption on Eve's attack discussed in Section~\ref{sec:channel-assumptions}. We have also supposed that Eve's system is a harmonic oscillator. Even though the mean photon number of the average input state in all example cases that we consider in what follows is many orders of magnitude smaller than $P=10^{7}$ and the actual physical channel being employed is a pure-loss channel, we can still suppose that the mean energy of the eavesdropper's states is extremely large (way beyond what an eavesdropper might reasonably employ in an attack) and we find that the performance of the discrete-modulation protocols approaches that of the Gaussian-modulation protocol relatively quickly as the constellation size $m^2$ increases. One could choose an even more conservative value for $P$, higher than what we have taken. However, in our numerics, we have found the same qualitative behavior: that the performance of the discrete-modulation protocol rapidly approaches that of the Gaussian-modulation protocol as the constellation size $m^2$ increases. 

With all these ingredients in hand, we can now numerically evaluate \eqref{eq:key_rate} to obtain a lower bound on the secret-key rate for a lossy thermal bosonic channel with transmissivity~$\eta$ and thermal noise as $N_B$. We take the reconciliation efficiency $\beta=0.95$. Note that the key rates obtained from \eqref{eq:key_rate} have a dependence on the mean photon number $N_S$ of the input state. Thus, to obtain tight lower bounds on the secret-key rate, we also optimize over $N_S$.

In Figure~\ref{fig:plots1}, we present the achievable key rates obtained from discrete-modulation protocol with homodyne detection if the underlying channel is a pure-loss channel. In
Figures~\ref{fig:plots2}, and \ref{fig:plots3}, we present the achievable key rates from discrete-modulation protocol with heterodyne detection if the underlying channel is a thermal channel. We plot lower bounds for various values of $m^2$ and  compare the obtained lower bounds with the Gaussian-modulation protocol. It is clear that the secure key rate of the discrete-modulation protocol increases as the constellation size $m^2$ increases. 

As explained before, to obtain the lower bound on the rates for pure-loss channel and thermal channel, we optimize over the mean photon number $N_S$. For the Gaussian-modulation protocol, we find that the optimal variance for secret-key rates decreases with the increase in loss. Now, the main idea behind the technique presented in this paper is to discretize the Gaussian probability distribution by a finite constellation of size $m^2$ and calculate the error introduced due to this approximation. We find that as the variance of the Gaussian modulation increases, the number of constellation points required to approximate the distribution to an $\varepsilon$ error increases. Therefore, for low losses, this technique requires a large number $m^2$ of constellation points to closely match the secret key-rates obtained with Gaussian modulation.

A consequence of the aforementioned reasoning is that, with our approach, the lower bound on the secret-key rate does not tend to $\log m^2$ in the limit as $\eta \rightarrow 1$. Certainly, in this limit, the Holevo information with Eve tends to zero, and the key rate is then given as $K\geq I(X;Y)-f(\varepsilon,N_S^{\prime})$. We numerically observe that the Shannon mutual information of Alice and Bob saturates towards $\log m^2$ with the increase in variance; however, the approximation error $f(\varepsilon,N_S^{\prime})$ increases with the increase in variance. Due to this trade-off, our technique does not achieve the ideal rate of $\log m^2$ rate in the low-loss and low-noise limit. 

It is possible (and likely) that our rate lower bounds can be improved by other constellation choices or other proof techniques. However, even with our proof, requiring a pair of electro-optic (phase and amplitude) modulators to generate a $90 \times 90$ size constellation size is much more practical and less demanding compared to asking that we modulate a pulse with a complex amplitude to an extremely high floating point accuracy, which a Gaussian modulation would need.

We should also point out that dark counts in detectors can be modeled as thermal noise for lossy thermal bosonic channels \cite{Rozpdek2018}. Thus, our numerics are also  applicable to protocols with imperfect detectors modeled in this way. 



We note here that we have included in the arXiv posting of this paper the Mathematica files used to perform the numerical calculation of the key rates and to generate the figures.

\section{Conclusion}\label{sec:conclusion}

In this paper, we have addressed an open problem in continuous-variable quantum key distribution (CV-QKD), by establishing a security proof for discrete-modulation protocols. Even though many experiments have been performed on discrete-modulation CV-QKD with multiple constellation points (see, e.g., \cite{Hirano2017,ZD17,Li:18}), no security proofs have been available for them, and the expected key-rate calculations previously reported based on measured homodyne statistics have been based on assuming Gaussian attacks, which are not known to be optimal for discrete-modulation CV-QKD. We have introduced a discrete-modulation protocol and then obtained rigorous lower bounds on the secret-key rates, secure against physically reasonable collective attacks in the asymptotic key-length regime. The approach that we have used works well in the high-loss regime, with the secure key rates being close to the secure key rates achievable with a Gaussian-modulation protocol. Another prominent feature of our approach is that with the increase in the size $m^2$ of the constellation, the lower bound on the secret-key rate approaches the key rate for the Gaussian-modulation protocol. This result demonstrates that we need not consider the full continuum of the Gaussian distribution in order to obtain key rates achievable with a Gaussian modulation, and we do not need to rely on Gaussian modulation for security proofs of discrete-modulation CV-QKD protocols.

Going forward from here, it is a pressing open question to determine security proofs for discrete-modulation CV-QKD protocols in the non-asymptotic, or finite key-length, regime. It would also be ideal to improve the bound from Proposition~\ref{prop:estimategaussian} to reduce or eliminate its dependence on the parameters $c_1$, $a$, and $K$. 
In this context, it might be possible to utilize the results presented in Appendix~\ref{sec:appendix-alternative}, but as mentioned previously, this approach is numerically intensive.


\begin{acknowledgments}
We are grateful to Hari Krovi for discussions about CV-QKD with discrete modulation. We are especially grateful to Anthony Leverrier for his critical reading of the initial version of our paper, for pointing out the need for more details of the channel estimation procedure, and for the idea behind Eq.~\eqref{eqn:isometry-define}. We thank Jeffrey H.~Shapiro, Xiang-Bin Wang, and Cosmo Lupo for helpful feedback on our manuscript. This work was supported by the Office of Naval Research program Communications and Networking with Quantum Operationally-Secure Technology for Maritime Deployment (CONQUEST): Raytheon BBN Technologies prime contract number N00014-16-C2069, under subcontracts to Louisiana State University and University of Arizona.
\end{acknowledgments}\vspace{1.5em}




\bibliography{main}

\onecolumngrid

\begin{appendices}

\section{Proof of Proposition~\ref{prop:estimategaussian}}

\label{app:estimategaussian}

In this appendix, we provide a proof of Proposition~\ref{prop:estimategaussian}. We first restate it here for convenience.

\begin{proposition}\label{prop:estimategaussian1}
Let $\overline{\rho}= \sum_x r_X(x) \op{\alpha_x}$, where
\begin{align}
\alpha_x & =\frac{q_A+ip_A}{\sqrt{2}}, \\
r_X(x) & = r_{Q_A}(q_A)\, r_{P_A}(p_A), \\
\theta_{N_S} & = \int dx\, r^G_X(x)\op{\alpha_x},
\end{align}
and $r^G(x)$ is the $P$-function of a thermal state with mean photon number $N_S$. If  $\sqrt{\chi^2(\overline{\rho},\theta(N_S))} \leq \varepsilon^2$, and Eve's attacks fulfill the constraints in Section~\ref{sec:channel-assumptions},
then
\begin{align}
\gamma_{11} &= \gamma_{11}^G, \label{app:eq:11-equality1}\\
|\gamma_{22}-\gamma_{22}^G| &\leq \varepsilon_1,\label{app:eq:12-inequality1}\\
|\gamma_{12}-\gamma_{12}^G| &\leq \varepsilon_2,\label{app:eq13-inequality1}
\end{align}
where 
\begin{align}
       \varepsilon_1 & = \varepsilon\cdot\left(1+c_1\right)\cdot\sqrt{\mathbb{E}\left[(Q_B- \mathbb{E}[Q_B])^4\right]},\\
\end{align}
for some constant $c_1>0$ and
\begin{multline}
\varepsilon_2 = \sum_{k=0}^{2m-2}\sum_{l=2m}^{K}\mu_{kl}\left|\eta^G(q_A,k+1)\left(\eta^G(p_A,l)-\eta(p_A,l)\right)\right|\\+\sum_{k=2m-1}^{K}\sum_{l=0}^{2m-1}\mu_{kl}\left|\eta^G(p_A,l)\left(\eta^G(q_A,k+1)-\eta(q_A,k+1)\right)\right|\\+\sum_{k=2m-1,l=2m}^{K}\mu_{kl}\left|\eta^G(p_A,l)\eta^G(q_A,k+1)-\eta(p_A,l)\eta(q_A,k+1)\right|,
\end{multline}
where $\mu_{kl}$ is an arbitrary function for $k\leq 2m-2,\ l\leq2m-1$ and is equal to $\exp({-a(k+l)})$ otherwise. We also have
\begin{align}
    \eta^G(q_A,k) & = \mathbb{E}_{r^G_{Q_A}}[Q_A^{k}],\\
    \eta(q_A,k)  & = \mathbb{E}_{r_{Q_A}}[Q_A^{k}],\\
    \eta^G(p_A,k) & = \mathbb{E}_{r^G_{P_A}}[P_A^{k}],\\
    \eta(p_A,k) & = \mathbb{E}_{r_{P_A}}[P_A^{k}].
\end{align}

\end{proposition}

\begin{proof}
For simplicity, we prove the claim under the assumption that all random variables have zero mean, and we note that it can be generalized by adopting a shift of the variables involved in the proof.

To prove the equality in \eqref{app:eq:11-equality1}, consider the following: $\gamma_{11}$ is equal to the variance of the position quadrature that is encoded by Alice during the preparation procedure. Since we are using the Gauss-Hermite  approximation  of the Gaussian for the encoding, for which the lower moments match those of the Gaussian distribution, it follows that $\gamma_{11}=\gamma_{11}^G$. 
To prove the inequality in \eqref{app:eq:12-inequality1}, observe that 
\begin{equation}
    \left\|\overline{\rho}-\theta(N_S)\right\|_1 \leq  
    \sqrt{\chi^2(\overline{\rho},\theta(N_S))}
    \leq \varepsilon^2 ,  \label{eq:distance_gaussian}
    \end{equation}
implies that    
\begin{equation}
    \left\|\mathcal{N}(\overline{\rho})-\mathcal{N}(\theta(N_S))\right\|_1 \leq \varepsilon^2.
    \label{eq:DP-TD-after-ch}
\end{equation}
The inequality in \eqref{eq:DP-TD-after-ch} follows from data processing.

Now let us define
\begin{align}
    r_{Q_B}(q_B)\equiv \int\int dx\ dp_B\  W^{\mathcal{N},\alpha_x}(q_B,p_B)r_X(x),\\
    r^G_{Q_B}(q_B)\equiv  \int\int dx\ dp_B\  W^{\mathcal{N},\alpha_x}(q_B,p_B)r^G_X(x),
\end{align}
where 
$W^{\mathcal{N},\alpha_x}$ is the associated Wigner function for the state resulting from transmitting a coherent state over the channel $\mathcal{N}$.
Let 
$r_{Q_B|X}(q_B|x) \equiv  \int dp_B\, W^{\mathcal{N},\alpha_x}(q_B,p_B)$ be the probability distribution obtained over the position quadrature when the coherent state $\alpha_x$ is sent over a channel $\mathcal{N}$. 

Then we have the following: 
\begin{align}
    \int dq_B \left\vert r_{Q_B}(q_B)-r^G_{Q_B}(q_B)\right\vert 
    &
    \leq \|\mathcal{N}(\bar{\rho})-\mathcal{N}(\theta(N_S))\|_1 \leq \varepsilon^2,
\end{align}
which is a consequence of monotonicity of trace distance and \eqref{eq:distance_gaussian}.


We obtain the following:%
\begin{align}
& |\gamma_{22}^G - \gamma_{22}| = \left\vert \int dq_B\ r^G_{Q_B}(q_B)\ \left\vert q_B\right\vert ^{2}-\int
dq_B\ r_{Q_B}(q_B)\ \left\vert q_B\right\vert ^{2}\right\vert \\ 
& =\left\vert \int dq_B\ \left[  r^G_{Q_B}(q_B)-r_{Q_B}(q_B)\right]  \ \left\vert
q_B\right\vert ^{2}\right\vert \\
& =\left\vert \int dq_B\ \left[  \sqrt{r^G_{Q_B}(q_B)}\sqrt{r^G_{Q_B}(q_B)}-\sqrt{r^G_{Q_B}%
(q_B)}\sqrt{r_{Q_B}(q_B)}+\sqrt{r^G_{Q_B}(q_B)}\sqrt{r_{Q_B}(q_B)}-\sqrt{r_{Q_B}(q_B)}\sqrt
{r_{Q_B}(q_B)}\right]  \ \left\vert q_B\right\vert ^{2}\right\vert \\
& \leq\int dq_B\ \left\vert \sqrt{r^G_{Q_B}(q_B)}-\sqrt{r_{Q_B}(q_B)}\right\vert
\ \sqrt{r^G_{Q_B}(q_B)}\left\vert q_B\right\vert ^{2}+\int dq_B\ \left\vert \sqrt
{r^G_{Q_B}(q_B)}-\sqrt{r_{Q_B}(q_B)}\right\vert \ \sqrt{r_{Q_B}(q_B)}\left\vert q_B\right\vert
^{2}\\
& \leq\sqrt{\int dq_B\ \left\vert \sqrt{r^G_{Q_B}(q_B)}-\sqrt{r_{Q_B}(q_B)}\right\vert
^{2}\ \int dq_B\ r^G_{Q_B}(q_B)\left\vert q_B\right\vert ^{4}}\nonumber\\
& \qquad+\sqrt{\int dq_B\ \left\vert \sqrt{r^G_{Q_B}(q_B)}-\sqrt{r_{Q_B}(q_B)}\right\vert
^{2}\ \int dq_B\ r_{Q_B}(q_B)\left\vert q_B\right\vert ^{4}}\\
& =\sqrt{\int dq_B\ \left\vert \sqrt{r^G_{Q_B}(q_B)}-\sqrt{r_{Q_B}(q_B)}\right\vert ^{2}%
}\left(  \sqrt{\int dq_B\ r^G_{Q_B}(q_B)\left\vert q_B\right\vert ^{4}}+\sqrt{\int
dq_B\ r_{Q_B}(q_B)\left\vert q_B\right\vert ^{4}}\right)  .
\end{align}
Now using that $\sqrt{\int dq_B \left\vert \sqrt{r^G_{Q_B}(q_B)}-\sqrt{r_{Q_B}%
(q_B)}\right\vert ^{2}}$ is the Hellinger divergence and less than the square root of the total variation distance $\int dq_B \left\vert r^G_{Q_B}(q_B)-r_{Q_B}
(q_B)\right\vert $, we obtain that%
\begin{equation}
\left\vert \int dq_B\ r^G_{Q_B}(q_B)\ \left\vert q_B\right\vert ^{2}-\int dq_B\ r_{Q_B}%
(q_B)\ \left\vert q_B\right\vert ^{2}\right\vert \leq\varepsilon\cdot\left(
\sqrt{\int dq_B\ r^G_{Q_B}(q_B)\left\vert q_B\right\vert ^{4}}+\sqrt{\int dq_B\ r_{Q_B}%
(q_B)\left\vert q_B\right\vert ^{4}}\right)  .
\end{equation}
To bound the second term we invoke the assumption that the photon number variance of the channel output is bounded. Therefore,
\begin{align}
    \operatorname{Tr}(\hat{n}^2\rho)&= \operatorname{Tr}\left((\hat{q}_B^2+\hat{p}_B^2-1)^2 \rho\right)\\
    &= \operatorname{Tr}\left((\hat{q}_B^4 +\hat{p}_B^4+1-2\hat{q}_B^2-2\hat{p}_B^2+2\hat{q}_B^2\hat{p}_B^2)\rho\right) < \infty,
\end{align}
where $\rho= \mathcal{N}\left(\theta(N_S)\right)$. We thus conclude that $\sqrt{\int dq_B\ r^G_{Q_B}%
(q_B)\left\vert q_B\right\vert ^{4}}$ is also bounded, so that  $\sqrt{\int dq_B\ r^G_{Q_B}
(q_B)\left\vert q_B\right\vert ^{4}}\leq c_1\cdot \sqrt{\int dq_B\ r_{Q_B}
(q_B)\left\vert q_B\right\vert ^{4}}$, for some constant $c_1 >0$.

To prove the inequality in \eqref{app:eq13-inequality1}, observe the following:
\begin{align}
    \gamma_{12}&= \int \int dq_A\ dp_A\ r_{Q_A}(q_A)\ r_{P_A}(p_A)q_A\int \int dp_B\ dq_B\ r_{Q_B,P_B|Q_A,P_A}(q_B\  p_B|q_A\ p_A)q_B.\\
    &= \int \int dq_A\ dp_A\ r_{Q_A}(q_A)\ r_{P_A}(p_A)q_A\ \int dq_B \ r_{Q_B|Q_A,P_A}(q_B|q_A\ p_A)q_B.
\end{align}
Let us define
\begin{equation}
    \mu(q_A,p_A) \equiv \int dq_B \ q_B \ r_{Q_B|Q_AP_A}(q_B|q_Ap_A).
\end{equation}
This implies that
\begin{equation}
    \gamma_{12}= \int \int dq_A\ dp_A\ r_{Q_A}(q_A)\ r_{P_A}(p_A)\ q_A\  \mu(q_A,p_A).
\end{equation}
Similarly, we can define 
\begin{equation}
    \gamma_{12}^G=\int \int dq_A\ dp_A\ r^G_{Q_A}(q_A)\ r_{P_A}^G(p_A)\ q_A\  \mu(q_A,p_A).
\end{equation}
This implies 
\begin{align}
    |\gamma_{12}-\gamma_{12}^G|=
    \left|\int dq_A\ dp_A\  \mu(q_A,p_A)\left(r^G(p_A)\ r^G(q_A)q_A-r_{P_A}(p_A)\ r_{Q_A}(q_A)q_A\right)\right|.
\end{align}
Now let us suppose that it is possible to expand $\mu(q_A,p_A)$ as a polynomial in $q_A$ and $p_A$, as mentioned in the assumptions from Section~\ref{sec:channel-assumptions}. That is, 
\begin{equation}\label{eqn:mu-expression}
    \mu(q_A,p_A)=\sum_{k=0}^{K}\sum_{l=0}^{K} \mu_{k,l}q_A^kp_A^l.
\end{equation}
This assumption implies that the mean value of Bob's position-quadrature measurement result when a coherent state $\op{\alpha(q_A,p_A)}$ is transmitted through an unknown channel $\mathcal{N}$ is no more than polynomial in $q_A$ and $p_A$. 

With this assumption, we obtain the following: 
\begin{align} \label{eq:difference}
    |\gamma_{12}-\gamma_{12}^G|=\left|\sum_{k,l}\mu_{k,l}\int \int dq_A \ dp_A\left(r^G_{P_A}(p_A)\ p_A^l\ r ^G_{Q_A}(q_A)\ q_A^{k+1}-r_{P_A}(p_A)\ p_A^l\ r_{Q_A}(q_A)\ q_A^{k+1}\right)\right|
\end{align}

Let us now define the following:
\begin{align}
\eta^G(p_A,l) & \equiv \int dp_A\ r_{P_A}^G(p_A)\ p_A^l,\\
\eta^G(q_A,l) & \equiv \int dq_A\ r_{Q_A}^G(q_A)\ q_A^l,\\
\eta(p_A,l) & \equiv \int dp_A\ r_{P_A}(p_A)\ p_A^l,\\
\eta(q_A,l) & \equiv \int dq_A\ r_{Q_A}(p_A)\ q_A^l.
\end{align}
This implies that
\begin{equation}
    |\gamma_{12}-\gamma_{12}^G|\leq \sum_{k,l}\mu_{k,l}\left|\eta^G(p_A,l)\eta^G(q_A,k+1)-\eta(p_A,l)\eta(q_A,k+1)\right|
\end{equation}

Now, we know that the first $2m-1$ moments of the Gauss-Hermite distribution are equal to the first $2m-1$ moments of the Gaussian distribution. With that, we conclude the following upper bound:
\begin{multline}
    |\gamma_{12}-\gamma_{12}^G|\leq \sum_{k=0}^{2m-2}\sum_{l=2m}^{K}\mu_{kl}\left|\eta^G(q_A,k+1)\left(\eta^G(p_A,l)-\eta(p_A,l)\right)\right|\\+\sum_{k=2m-1}^{K}\sum_{l=0}^{2m-1}\mu_{kl}\left|\eta^G(p_A,l)\left(\eta^G(q_A,k+1)-\eta(q_A,k+1)\right)\right|\\+\sum_{k=2m-1,l=2m}^{K}\mu_{kl}\left|\eta^G(p_A,l)\eta^G(q_A,k+1)-\eta(p_A,l)\eta(q_A,k+1)\right|.
\end{multline}
This concludes the proof.
\end{proof}

\section{Channel symmetrization}

\label{appendix_sym}

We now show that by performing a discrete phase symmetrization  in Steps~2-3 of the key distribution protocol from Section~\ref{sec:protocol}, it is possible to simplify the form of the covariance matrix of the state that Alice and Bob share at the end of the EB protocol to a symmetrized form. 

Let $\mathcal{N}_{A\rightarrow B}$ be a single-mode bosonic channel. Alice and
Bob can make this channel phase covariant by applying a random phase rotation
and its inverse at the channel input and output, respectively, resulting in
the following symmetrized channel:
\begin{equation}
\overline{\mathcal{N}}_{A\rightarrow B}(\rho_{A})=\frac{1}{4}\sum_{k=0}%
^{3}e^{i\hat{n}_{B}\pi k/2}\mathcal{N}_{A\rightarrow B}(e^{-i\hat{n}_{A}\pi
k/2}\rho_{A}e^{i\hat{n}_{A}\pi k/2})e^{-i\hat{n}_{B}\pi k/2}.
\end{equation}
If the state input to the phase randomized channel is one share of a two-mode
squeezed vacuum $\psi(\bar{n})_{RA} = | \psi(\bar{n}) \rangle \langle  \psi(\bar{n}) |_{RA}$, defined from%
\begin{equation}
|\psi(\bar{n})\rangle_{RA}\equiv\frac{1}{\sqrt{\bar{n}+1}}\sum_{n=0}^{\infty
}\sqrt{\left(  \frac{\bar{n}}{\bar{n}+1}\right)  ^{n}}|n\rangle_{R}%
\otimes|n\rangle_A,
\end{equation}
then it follows that%
\begin{align}
\overline{\mathcal{N}}_{A\rightarrow B}(\psi(\bar{n})_{RA})=
\frac{1}{4}\sum_{k=0}^{3}\left(  e^{-i\hat{n}_{R}\pi k/2}\otimes e^{i\hat
{n}_{B}\pi k/2}\right)  \mathcal{N}_{A\rightarrow B}(\psi(\bar{n}%
)_{RA})\left(  e^{i\hat{n}_{R}\pi k/2}\otimes e^{-i\hat{n}_{B}\pi k/2}\right)
,
\end{align}
where we have applied the fact that%
\begin{equation}
e^{-i\hat{n}_{A}\pi k/2}|\psi(\bar{n})\rangle_{RA}=e^{-i\hat{n}_{R}\pi
k/2}|\psi(\bar{n})\rangle_{RA}.
\end{equation}
We would now like to determine the covariance matrix elements of the
phase-randomized state $\tau_{RB}\equiv\overline{\mathcal{N}}_{A\rightarrow
B}(\psi(\bar{n})_{RA})$:%
\begin{equation}%
\begin{bmatrix}
2\left\langle \hat{x}_{R}^{2}\right\rangle _{\tau} & \left\langle \left\{
\hat{x}_{R},\hat{p}_{R}\right\}  \right\rangle _{\tau} & \left\langle \left\{
\hat{x}_{R},\hat{x}_{B}\right\}  \right\rangle _{\tau} & \left\langle \left\{
\hat{x}_{R},\hat{p}_{B}\right\}  \right\rangle _{\tau}\\
\left\langle \left\{  \hat{x}_{R},\hat{p}_{R}\right\}  \right\rangle _{\tau} &
2\left\langle \hat{p}_{R}^{2}\right\rangle _{\tau} & \left\langle \left\{
\hat{p}_{R},\hat{x}_{B}\right\}  \right\rangle _{\tau} & \left\langle \left\{
\hat{p}_{R},\hat{p}_{B}\right\}  \right\rangle _{\tau}\\
\left\langle \left\{  \hat{x}_{R},\hat{x}_{B}\right\}  \right\rangle _{\tau} &
\left\langle \left\{  \hat{p}_{R},\hat{x}_{B}\right\}  \right\rangle _{\tau} &
2\left\langle \hat{x}_{B}^{2}\right\rangle _{\tau} & \left\langle \left\{
\hat{x}_{B},\hat{p}_{B}\right\}  \right\rangle _{\tau}\\
\left\langle \left\{  \hat{x}_{R},\hat{p}_{B}\right\}  \right\rangle _{\tau} &
\left\langle \left\{  \hat{p}_{R},\hat{p}_{B}\right\}  \right\rangle _{\tau} &
\left\langle \left\{  \hat{x}_{B},\hat{p}_{B}\right\}  \right\rangle _{\tau} &
2\left\langle \hat{p}_{B}^{2}\right\rangle _{\tau}%
\end{bmatrix}
,
\end{equation}
where we assume for simplicity that $\tau_{RB}$ has zero mean, but we note here that the more general case can be incorporated by a shift.
Given an initial covariance matrix with elements%
\begin{equation}
\sigma=%
\begin{bmatrix}
\sigma_{11} & \sigma_{12} & \sigma_{13} & \sigma_{14}\\
\sigma_{12} & \sigma_{22} & \sigma_{23} & \sigma_{24}\\
\sigma_{13} & \sigma_{23} & \sigma_{33} & \sigma_{34}\\
\sigma_{14} & \sigma_{24} & \sigma_{34} & \sigma_{44}%
\end{bmatrix}
,
\end{equation}
the phase rotation $e^{-i\hat{n}_{R}\phi}\otimes e^{i\hat{n}_{B}\phi}%
$\ corresponds to the following symplectic transformation%
\begin{equation}
X(\phi)=%
\begin{bmatrix}
\cos(\phi) & \sin(\phi) & 0 & 0\\
-\sin(\phi) & \cos(\phi) & 0 & 0\\
0 & 0 & \cos(\phi) & -\sin(\phi)\\
0 & 0 & \sin(\phi) & \cos(\phi)
\end{bmatrix}
.
\end{equation}
So then calculating covariance matrix for the phase randomized state, it is
given by%
\begin{equation}
\frac{1}{4}\sum_{k=0}^{3}X(\pi k/2)\sigma X^{T}\!(\pi k/2),
\end{equation}
and we find that it is equal to%
\begin{equation}
\frac{1}{2}%
\begin{bmatrix}
\sigma_{11}+\sigma_{22} & 0 & \sigma_{13}-\sigma_{24} & \sigma_{14}%
+\sigma_{23}\\
0 & \sigma_{11}+\sigma_{22} & \sigma_{14}+\sigma_{23} & -\left(  \sigma
_{13}-\sigma_{24}\right)  \\
\sigma_{13}-\sigma_{24} & \sigma_{14}+\sigma_{23} & \sigma_{33}+\sigma_{44} &
0\\
\sigma_{14}+\sigma_{23} & -\left(  \sigma_{13}-\sigma_{24}\right)   & 0 &
\sigma_{33}+\sigma_{44}%
\end{bmatrix}
.
\end{equation}
The latter has the following form:%
\begin{equation}%
\begin{bmatrix}
a & 0 & c_{2} & c_{1}\\
0 & a & c_{1} & -c_{2}\\
c_{2} & c_{1} & b & 0\\
c_{1} & -c_{2} & 0 & b
\end{bmatrix}
,
\end{equation}
for $a,b\geq1$ and $c_{1},c_{2}\in\mathbb{R}$. We can write this in the form
of Eq.~(D34) in \cite{L15} by setting $c_{1}=z\sin(\theta)$ and $c_{2}=z\cos
(\theta)$, so that $z=\sqrt{c_{1}^{2}+c_{2}^{2}}$ and $\theta=\arctan
(c_{1}/c_{2})$, so that the form becomes%
\begin{equation}%
\begin{bmatrix}
a & 0 & z\cos(\theta) & z\sin(\theta)\\
0 & a & z\sin(\theta) & -z\cos(\theta)\\
z\cos(\theta) & z\sin(\theta) & b & 0\\
z\sin(\theta) & -z\cos(\theta) & 0 & b
\end{bmatrix} = 
\begin{bmatrix}
a \mathbb{I}_2  & z\, R(\theta)  \\
z\, R(\theta) & b \mathbb{I}_2
\end{bmatrix}.
\end{equation}
This completes the symmetrization of the covariance matrix due to the discrete phase randomization. 

Ideally, we would estimate all the elements of the covariance matrix in the channel estimation step of the protocol. However, it is much simpler to estimate only the parameters $a$, $b$, and $z\cos(\theta)$, and assume instead that  the  covariance matrix has the following form:
\begin{equation}%
\begin{bmatrix}
a & 0 & z\cos(\theta) & 0\\
0 & a & 0 & -z\cos(\theta)\\
z\cos(\theta) & 0 & b & 0\\
0 & -z\cos(\theta) & 0 & b
\end{bmatrix}
.
\end{equation}
That is, we ignore all correlations between position and momentum quadratures. The effect of doing so is to underestimate the correlations that are present in the state $\tau_{RB}$. Therefore, this intuitively means we overestimate Eve's Holevo information during the channel estimation phase, and exhaustive numerical checks confirm that Eve's Holevo information is larger when replacing $z$ with $z\cos(\theta)$ (as was reported in \cite{L15}). Since we overestimate the Holevo information, the security of the protocol is not compromised, but it is only the final key rate that is potentially reduced. For additional discussion, see Appendix D of~\cite{L15}.

\section{Alternative approach for bounding the parameters of a hypothetical Gaussian-modulated protocol}
In this section, we outline a method to remove the parameters $c_1$ and $c_2$ from the security proof of discrete-modulation protocols. 

Let $\overline{\rho}$ be the averaged state that Alice sends to Bob in the  discrete-modulation protocol, and let the thermal state $\theta(N_S)$ be the averaged state that Alice sends to Bob in Gaussian-modulation protocol. Let $\mathcal{S}$ be a set of isometries $\mathcal{U}_{A\rightarrow BE}$ that Eve implements and which agrees with the statistics that are collected by Alice and Bob, and fullfills the criteria given in Section~\ref{sec:channel-assumptions}. Then, we want to obtain an upper bound on the Holevo information $
\sup_{\mathcal{U}\in \mathcal{S}}\chi(B;E)_{\mathcal{U}(\overline{\rho})} $.
Let $\mathcal{W}=\mathrm{Span}\left\{\ket{\alpha_1}, \ket{\alpha_2},\ldots \ket{\alpha_{m^2}}\right\}$ be a finite-dimensional vector space, and is spanned by a basis having $m^2$ elements. We want to construct an orthonormal basis for this space, and this can be done by using Gram-Schmidt orthogonalization method. Let 
\begin{equation}
    \ket{\phi_i'} = \ket{\alpha_i} - \sum_{j=1}^{i-1} \braket{\phi_j'}{\alpha_i} \ket{\phi_j'},
\end{equation}
and define $\ket{\phi_i} = \frac{\ket{\phi'_i}}{\Vert\phi'_i\Vert}$. 
We can then construct the following orthonormal basis $\left\{\ket{\phi_1},\ket{\phi_2}\ldots \ket{\phi}_{m^2}\right\}$ for $\mathcal{W}$. Now, let us define the following projector on $\mathcal{W}$: 
\begin{equation}
\Pi_W = \sum_{i=1}^{m^2} \op{\phi_i}.
\end{equation}
It is easy to see that 
\begin{equation}
\Pi_W\overline{\rho} \Pi_W = \overline{\rho}. 
\end{equation}
To each isometry $\mathcal{U}_{A'\rightarrow BE}\in \mathcal{S}$, we can define the following isometry $\overline{\mathcal{U}}$ such that
\begin{equation}
\operatorname{Tr}_{E'}\left[\overline{\mathcal{U}}_{A'\rightarrow BEE'}(\rho_{A'})\right] =\mathcal{U}_{A'\rightarrow BE}(\Pi_W \rho_{A'} \Pi_W) + \operatorname{Tr}\left[\left(1-\Pi_W\right) \rho_{A'}\right]\op{u}_{BE} \label{eqn:isometry-define},
\end{equation}
where $\ket{u}_{BE}$ is an arbitrary unit vector in $\mathcal{H}_{BE}$.
Then, $\operatorname{Tr}_{E'}\left[\mathcal{\overline{U}}_{A'\rightarrow BEE'}(\overline{\rho}_{A'})\right]= \mathcal{U}_{A'\rightarrow BE}(\Pi_W \overline{\rho}_{A'} \Pi_W)$. We also have that
\begin{equation}
\chi(B;EE')_{\mathcal{\overline{U}}(\bar{\rho})}\geq \chi(B;E)_{\mathcal{\overline{U}}(\overline{\rho})} = \chi(B;E)_{\mathcal{U}(\overline{\rho})}
\end{equation}
Then, using the continuity of Holevo information, we obtain
\begin{equation}\label{eq:holevo-info-alternative-method}
\chi(B;E)_{\overline{\mathcal{U}}(\overline{\rho})} \leq \chi(B;EE')_{\overline{\mathcal{U}}(\theta(N_S))}+ f(\varepsilon,N_S),
\end{equation}
where $f(\varepsilon,N_s)$ is defined in \eqref{eq:continuity_error}. Now we need to obtain an upper bound on $\chi(B;EE')_{\overline{\mathcal{U}}(\theta(N_S))}$. This Holevo information is calculated for a thermal state $\theta(N_S)$ sent over an isometric channel $\overline{\mathcal{U}}_{A'\rightarrow BEE'}$ in the set $\mathcal{S}$ and Bob performing homodyne or heterodyne measurement. For this, we obtain the parameters $\bar{\gamma}_{11}^G,\bar{\gamma}_{12}^G$ and $\bar{\gamma}_{22}^G$, which are defined analogously to \eqref{eq:DM-gam-11}--\eqref{eq:DM-gam-22}, with the initial random variable $Q_A$ replaced with Gaussian random variable with mean zero and variance equal to $N_S$. In the following proposition, we obtain bounds on the parameters $\bar{\gamma}_{11}^G,\bar{\gamma}_{12}^G$ and $\bar{\gamma}_{22}^G$ with respect to $\gamma_{11},\gamma_{12}$, and $\gamma_{22}$ observed in discrete-modulation protocol.  
\label{sec:appendix-alternative}

\begin{proposition}
\label{prop:estimategaussian2}
Let $\overline{\rho}= \sum_x r_X(x) \op{\alpha_x}$, where
\begin{align}
\alpha_x & =\frac{q_{A_s}+ip_{A_t}}{\sqrt{2}}, \\
r_X(x) & = r_{Q_{A}}(q_{A_s})\, r_{P_{A}}(p_{A_t}), \\
\theta_{N_S} & = \int dx\, r^G(x)\op{\alpha_x},
\end{align}
where $r^G(x)$ is the $P$-function for a thermal state with mean photon number $N_S$, and $s,t \in \left\{1\cdots m\right\}$. If  $\sqrt{\chi^2(\overline{\rho},\theta(N_S))}  \leq \varepsilon^2$ and Eve's attack    $\, \mathcal{U}_{A'\to B'E'}$ fulfills the constraints in Section~\ref{sec:channel-assumptions},
then,
\begin{align}
\bar{\gamma}^G_{11} &= \gamma_{11},\label{eqn:first-variable}\\
\bar{\gamma}_{22}^G &\leq \gamma_{22} + \gamma_{22}\left\Vert\overline{\rho}^{(-\frac{1}{2})}\right\Vert_{\infty}^2 \varepsilon +\varepsilon \operatorname{Tr}\left[\op{u} \hat{q}^2\right],
\\
\bar{\gamma}_{12}^G &\geq z\gamma_{12}+\int dq_A r^G(q_A) q_A \int dp_A r^G(p_A) \operatorname{Tr}\left[\left(\mathbb{I}-\Pi_W\right)\op{\alpha(q_A,p_A)}\right]\operatorname{Tr}\left[\op{u}\hat{q}\right], 
\end{align}
where $\Pi_W = \sum_{i=1}^{m^2} \op{\phi_i}$, $\ket{u}$ represents an arbitrary unit vector and $\ket{\phi_{i}}$ with $i \in \left\{1,\cdots m^2\right\}$, forms an orthonormal basis for $\mathcal{W}=\mathrm{Span}\left\{\ket{\alpha_1}, \ket{\alpha_2},\ldots \ket{\alpha_{m^2}}\right\}$, and 
\begin{equation}
z = \min_{i,j}\left|\frac{\int dq_A\int dp_A r^G(q_A)r^G(p_A)q_Ad_{ij}(q_A,p_A)}{\sum_{s,t}q_{A_s} r(q_{A_s}) r(p_{A_t})b_{ij}(s,t)}\right|,
\end{equation}
with 
$d_{ij}(q_A,p_A)=\braket{\phi_i}{\alpha(q_A,p_A)}\braket{\alpha(q_A,p_A)}{\phi_j}$, and $b_{ij}(s,t)=\braket{\phi_i}{\alpha(q_{A_s},p_{A_t})}\braket{\alpha(q_{A_s},p_{A_t})}{\phi_j}$. 
\end{proposition}

\begin{proof}
First, consider $\bar{\gamma}_{22}^G$ defined as
\begin{equation}
\bar{\gamma}_{22}^G = \operatorname{Tr} \left[\hat{q}^2 \mathcal{N}(\Pi_W \theta(N_S)\Pi_W)\right] +\operatorname{Tr}\left[\left(1-\Pi_W\right)(\theta(N_S))\right] \operatorname{Tr}\left[\op{u} \hat{q}^2\right]. 
\end{equation}
Let us concentrate on the first term $\operatorname{Tr} \left[\hat{q}^2 \mathcal{N}(\Pi_W \theta(N_S)\Pi_W)\right] = \operatorname{Tr} \left[\hat{q}^2 \mathcal{N}(\sigma)\right] $, where $\sigma= \Pi_W \theta(N_S)\Pi_W$. 
Then, 
\begin{align}
    \operatorname{Tr}\left[\hat{q}^2 \mathcal{N}(\sigma)\right] 
    & = \operatorname{Tr}\left[\hat{q}^2 \mathcal{N}(\overline{\rho}^{(\frac{1}{2})}\overline{\rho}^{(-\frac{1}{2})}\sigma\overline{\rho}^{(-\frac{1}{2})}\overline{\rho}^{(\frac{1}{2})})\right] \label{eqs:alternateeq1}\\
    & \leq \operatorname{Tr}\left[\hat{q}^2 \mathcal{N}(\overline{\rho})\right]\Vert\overline{\rho}^{(-\frac{1}{2})}\sigma\overline{\rho}^{(-\frac{1}{2})}\Vert_{\infty}\label{eqs:alternateeq2}\\
    &\leq \gamma_{22}\, \Vert\overline{\rho}^{(-\frac{1}{2})}\sigma\overline{\rho}^{(-\frac{1}{2})}\Vert_{\infty}.
\end{align}
To obtain \eqref{eqs:alternateeq2} from \eqref{eqs:alternateeq1}, observe that 
\begin{align}
\rho^{-1/2} \sigma \rho^{-1/2} &\leq \|\rho^{-1/2} \sigma \rho^{-1/2}\|_{\infty} \ I\\
\implies \rho^{1/2}\rho^{-1/2} \sigma \rho^{-1/2}\rho^{1/2}& \leq \|\rho^{-1/2} \sigma \rho^{-1/2}\|_{\infty} \ \rho\\
\implies \operatorname{Tr}\left[\hat{q}^2\rho^{1/2}\rho^{-1/2} \sigma \rho^{-1/2}\rho^{1/2}\right] &\leq \|\rho^{-1/2} \sigma \rho^{-1/2}\|_{\infty} \operatorname{Tr}\left[\hat{q}^2\rho\right]
\end{align}
Then, from linearity of channel, the inequality follows. 

Now, we would like to obtain an upper bound on $\Vert\overline{\rho}^{(-\frac{1}{2})}\sigma\overline{\rho}^{(-\frac{1}{2})}\Vert_{\infty}$.
Let $\Delta=\sigma-\overline{\rho}$. Then, 
\begin{align}
  \Vert\overline{\rho}^{(-\frac{1}{2})}\sigma\overline{\rho}^{(-\frac{1}{2})}\Vert_{\infty}
  &= \Vert\overline{\rho}^{(-\frac{1}{2})}\left(\Delta +\overline{\rho}\right)\overline{\rho}^{(-\frac{1}{2})}\Vert_{\infty}\\
  & =\Vert\overline{\rho}^{(-\frac{1}{2})}\Delta \overline{\rho}^{(-\frac{1}{2})}+ \mathbb{I}\Vert_{\infty}\\
  &\leq \Vert\mathbb{I}\Vert_{\infty}+ \Vert\overline{\rho}^{(-\frac{1}{2})}\Delta \overline{\rho}^{(-\frac{1}{2})}\Vert_{\infty}\\
  &\leq 1 + \Vert\overline{\rho}^{(-\frac{1}{2})}\Vert_{\infty}^2\Vert\Delta\Vert_{\infty}\\
  &\leq 1 + \Vert\overline{\rho}^{(-\frac{1}{2})}\Vert_{\infty}^2\Vert\Delta\Vert_{1}.
\end{align}
Now, 
$\Vert\overline{\rho}-\theta(N_S)\Vert_1 \leq \varepsilon$. By data-processing we obtain
\begin{equation}
    \Vert \overline{\rho}-\sigma\Vert_1 \leq \varepsilon. 
\end{equation}
We thus obtain
\begin{align}
    \operatorname{Tr}\left[\hat{q}^2 \mathcal{N}(\sigma)\right] &\leq \gamma_{22} + \gamma_{22}\Vert\overline{\rho}^{(-\frac{1}{2})}\Vert_{\infty}^2\Vert\Delta\Vert_{1},\\
    &\leq \gamma_{22} + \gamma_{22}\Vert\overline{\rho}^{(-\frac{1}{2})}\Vert_{\infty}^2 \varepsilon.
\end{align}
Now, consider the following term: 
$
    \operatorname{Tr}\left[\left(\mathbb{I}-\Pi_W\right)\theta(N_S)\right]
$. We know that
\begin{align}
\|\overline{\rho}-\theta(N_S)\|_1 \leq \varepsilon
\end{align}
This implies, 
\begin{equation}
    \sup_{0\leq M\leq 1}\operatorname{Tr}\left[M\left(\theta(N_S)-\overline{\rho}\right)\right]\leq \varepsilon.
\end{equation}
Choosing $M = \mathbb{I}-\Pi_W$, we obtain, $
    \operatorname{Tr}\left[\left(\mathbb{I}-\Pi_W\right)\theta(N_S)\right] \leq \varepsilon$.
Then, 
\begin{equation}
\bar{\gamma}_{22}^G \leq \gamma_{22} + \gamma_{22}\Vert\overline{\rho}^{(-\frac{1}{2})}\Vert_{\infty}^2 \varepsilon +\varepsilon \operatorname{Tr}\left[\op{u} \hat{q}^2\right]. 
\end{equation}

Now, let us consider the parameter $\bar{\gamma}_{12}^G$. Numerical checks, similar to those stated in \cite{Leverrier1}, reveal that Holevo information is a monotonically decreasing function of this parameter. Since we want an upper bound on the Holevo information, we obtain a lower bound on this parameter. First, consider $\bar{\gamma}_{12}^{G,1}$ defined as follows:
\begin{align}
    \bar{\gamma}_{12}^{G,1} &= \int dq_A \ r^G(q_A) \ q_A \int dp_A r^G(p_A)\int dq_B \operatorname{Tr}\left[\mathcal{N}\left(\Pi_W\op{\alpha(q_A,p_A)}\Pi_W\right)\hat{q}_B\right]\\
    &= \int dq_A \ r^G(q_A) \ q_A \int dp_A r^G(p_A)\int dq_B \operatorname{Tr}\left[\mathcal{N}\left(\sum_{i,j}d_{ij}(q_A,p_A)\op{\phi_i}{\phi_j}\right)\hat{q}_B\right]\\
    &= \int dq_A \ r^G(q_A) \ q_A \int dp_A r^G(p_A) \sum_{i,j}d_{ij}(q_A,p_A)\int dq_B\operatorname{Tr}\left[\mathcal{N}\left(\op{\phi_i}{\phi_j}\right)\hat{q}_B\right]\\
   &= \sum_{i,j}\int dq_A \ r^G(q_A) \ q_A \int dp_A r^G(p_A) d_{ij}(q_A,p_A)\int dq_B\operatorname{Tr}\left[\mathcal{N}\left(\op{\phi_i}{\phi_j}\right)\hat{q}_B\right] \label{eqn:sum&integral}\\
   &= \sum_{i,j} f_{ij}\int dq_B\operatorname{Tr}\left[\mathcal{N}\left(\op{\phi_i}{\phi_j}\right)\hat{q}_B\right],
\end{align}
where $d_{ij}(q_A,p_A)=\braket{\phi_i}{\alpha(q_A,p_A)}\braket{\alpha(q_A,p_A)}{\phi_j}$, and $f_{ij}=\int dq_A\int dp_A r^G(q_A)r^G(p_A)q_Ad_{ij}(q_Ap_A)$. 
We can write $\gamma_{12}$ defined in \eqref{eqn:covarinace-paramter} as
\begin{align}
    \gamma_{12} &= \sum_{s,t} q_{A_s} r(q_{A_s}) r(p_{A_t})\operatorname{Tr}\left[\mathcal{N}\left(\op{\alpha(q_{A_s},p_{A_t})}\right)\hat{q}_B\right]\\&=\sum_{s,t} q_{A_s} r(q_{A_s}) r(p_{A_t})\operatorname{Tr}\left[\sum_{i,j}b_{ij}(s,t)\mathcal{N}\left(\op{\phi_i}{\phi_j}\right)\hat{q}_B\right]\\
    &=\sum_{i,j}\sum_{s,t} q_{A_s} r(q_{A_s}) r(p_{A_t})b_{ij}(s,t)\operatorname{Tr}\left[\mathcal{N}\left(\op{\phi_i}{\phi_j}\right)\hat{q}_B\right]\\
    &=\sum_{i,j}g_{ij}\operatorname{Tr}\left[\mathcal{N}\left(\op{\phi_i}{\phi_j}\right)\hat{q}_B\right],
\end{align}
where $b_{ij}(s,t)=\braket{\phi_i}{\alpha(q_{A_s},p_{A_t})}\braket{\alpha(q_{A_s},p_{A_t})}{\phi_j}$, and  $g_{ij}=\sum_{s,t}q_{A_s} r(q_{A_s}) r(p_{A_t})b_{ij}(s,t)$. We can then express $\bar{\gamma}_{12}^{G,1}$ in terms of $\bar\gamma_{12}$ as follows: 
\begin{align}
    \bar{\gamma}_{12}^{G,1}
    &= \sum_{i,j}\frac{f_{ij}}{g_{ij}}g_{ij}\operatorname{Tr}\left[\mathcal{N}\left(\op{\phi_i}{\phi_j}\right)\hat{q}_B\right]\\
    &\geq z \ \gamma_{12} \label{eq:lower-bound-covariance1},
\end{align}
where $z = \min_{i,j}\left|\frac{f_{ij}}{g_{ij}}\right|$. Let us now define $\bar{\gamma}_{12}^{G,2}$ as 
\begin{align} \label{eq:lower-bound-covariance2}
    \bar{\gamma}_{12}^{G,2} = \int dq_A r^G(q_A) q_A \int dp_A r^G(p_A) \operatorname{Tr}\left[\left(\mathbb{I}-\Pi_W\right)\op{\alpha(q_A,p_A)}\right]\operatorname{Tr}\left[\op{u}\hat{q}_B\right].
\end{align}
We can calculate the above term numerically. 

We can now combine \eqref{eq:lower-bound-covariance1} and \eqref{eq:lower-bound-covariance2} to obtain the following lower bound on $\bar{\gamma}_{12}^G= \bar{\gamma}_{12}^{G,1}+\bar{\gamma}_{12}^{G,2}$:
\begin{equation}
\bar{\gamma}_{12}^G\geq   z\gamma_{12}+\int dq_A r^G(q_A) q_A \int dp_A r^G(p_A) \operatorname{Tr}\left[\left(\mathbb{I}-\Pi_W\right)\op{\alpha(q_A,p_A)}\right)\operatorname{Tr}\left(\op{u}\hat{q}_B\right].
\end{equation}
This concludes the proof.
\end{proof}

\begin{corollary}
Let $\overline{\rho}= \sum_x r_X(x) \op{\alpha_x}$, where
\begin{align}
\alpha_x & =\frac{q_{A_s}+ip_{A_t}}{\sqrt{2}}, \\
r_X(x) & = r_{Q_{A}}(q_{A_s})\, r_{P_{A}}(p_{A_t}), \\
\theta_{N_S} & = \int dx\, r^G(x)\op{\alpha_x},
\end{align}
where $r^G(x)$ is the $P$-function for a thermal state with mean photon number $N_S$, and $s,t \in \left\{1\cdots m\right\}$. If  $\sqrt{\chi^2(\overline{\rho},\theta(N_S))}  \leq \varepsilon^2$ and Eve's attack    $\, \mathcal{U}_{A'\to B'E'}$ fulfills the constraints in Section~\ref{sec:channel-assumptions},
then,
\begin{align}
\bar{\gamma}^G_{11} &= \gamma_{11},\label{eqn:first-variable1}\\
\bar{\gamma}_{22}^G &\leq \gamma_{22} + \gamma_{22}\left\Vert\overline{\rho}^{(-\frac{1}{2})}\right\Vert_{\infty}^2 \varepsilon,
\\
\bar{\gamma}_{12}^G &\geq z\gamma_{12}, 
\end{align}
where $\Pi_W = \sum_{i=1}^{m^2} \op{\phi_i}$ and $\ket{\phi_{i}}$ with $i \in \left\{1,\cdots m^2\right\}$, forms an orthonormal basis for $\mathcal{W}=\mathrm{Span}\left\{\ket{\alpha_1}, \ket{\alpha_2},\ldots \ket{\alpha_{m^2}}\right\}$, and 
\begin{equation}
z = \min_{i,j}\left|\frac{\int dq_A\int dp_A r^G(q_A)r^G(p_A)q_Ad_{ij}(q_A,p_A)}{\sum_{s,t}q_{A_s} r(q_{A_s}) r(p_{A_t})b_{ij}(s,t)}\right|,
\end{equation}
with 
$d_{ij}(q_A,p_A)=\braket{\phi_i}{\alpha(q_A,p_A)}\braket{\alpha(q_A,p_A)}{\phi_j}$, and $b_{ij}(s,t)=\braket{\phi_i}{\alpha(q_{A_s},p_{A_t})}\braket{\alpha(q_{A_s},p_{A_t})}{\phi_j}$. 
\end{corollary}
\begin{proof}
In Proposition~\ref{prop:estimategaussian2}, choose $\ket{u}$ to be in the kernel of $\hat{q}$ (for example, a position-squeezed vacuum state that converges to a position eigenstate). 
\end{proof}

\bigskip
In Proposition~\ref{prop:estimategaussian2}, we obtain a lower bound on  $\bar{\gamma}_{12}^G$ and upper bound on $\bar{\gamma}_{22}^G$. Now, we can follow the steps stated in Section~\ref{sec:channel_estimation} to obtain an upper bound on $
\chi(B;E)_{\overline{\mathcal{U}}(\theta(N_S))}$. Consider a state $\sigma_{ABEE'} = \operatorname{Tr}\left[\overline{\mathcal{U}}_{A'\rightarrow BEE'} (\psi(\bar{n}))_{AA'}\right]$, where $\ket{\psi(\bar{n}}_{AA'}$ is TMSV as defined in \eqref{eq:TMSV}. Let the covariance matrix of $\sigma_{AB}$ be 
\begin{equation}
\begin{bmatrix}
\bar{\gamma}_{11}^{\operatorname{EB}} \, \mathbb{I}_2 & \bar{\gamma}_{12}^{\operatorname{EB}}\sigma_Z\\
\bar{\gamma}_{12}^{\operatorname{EB}}\sigma_Z & \bar{\gamma}_{22}^{\operatorname{EB}}\,  \mathbb{I}_2
\end{bmatrix}.
\label{eq:EB-cov-matrix1}
\end{equation}
We can use the ``PM to EB" mapping defined in Section~\ref{sec:est-parameters-hypothetical} to obtain bounds on parameters $\bar{\gamma}_{11}^{\operatorname{EB}}$, $\bar{\gamma}_{12}^{\operatorname{EB}}$, and $\bar{\gamma}_{22}^{\operatorname{EB}}$ of state $\sigma_{AB}$ from bounds on $\bar{\gamma}_{11}, \bar{\gamma}_{12}$ and $\bar{\gamma}_{22}$. Then invoke the Gaussian extremality theorem to state that the Holevo information $\chi(B;E)_{\overline{\mathcal{U}}(\theta(N_S))}$ is maximized by a Gaussian state with the covariance matrix given in \eqref{eq:EB-cov-matrix1}. By combining the upper bound obtained on $
\chi(B;E)_{\overline{\mathcal{U}}(\theta(N_S))}$ with \eqref{eq:holevo-info-alternative-method}, we obtain an upper bound on $\chi(B;E)_{\mathcal{U}(\overline{\rho})}$.

The method outlined above does succeed in obtaining a security proof for discrete-modulation protocols with no dependence on the parameters $c_1$ and $c_2$. However, as of now it seems that this method is numerically intensive.

\end{appendices}

\end{document}